\documentclass[preprint,prd,tightenlines,nofootinbib,eqsecnum,superscriptaddress]{revtex4-1}

\usepackage{amsmath}
\usepackage{amsfonts}
\usepackage{amssymb}
\usepackage{mathrsfs}
\usepackage{bm}
\usepackage{graphicx}
\usepackage{color}
\usepackage{array}
\usepackage{booktabs}
\usepackage{verbatim}
\usepackage{pagecolor}
\usepackage{hyperref}

\newcolumntype{C}{>{$}c<{$}}
\AtBeginDocument{
\heavyrulewidth=.08em
\lightrulewidth=.05em
\cmidrulewidth=.03em
\belowrulesep=.65ex
\belowbottomsep=0pt
\aboverulesep=.4ex
\abovetopsep=0pt
\cmidrulesep=\doublerulesep
\cmidrulekern=.5em
\defaultaddspace=.5em
}

\definecolor{CiteColor}{rgb}{0,0,0.35}
\definecolor{URLColor}{rgb}{0,0,0.35}
\hypersetup{colorlinks,citecolor=CiteColor,urlcolor=URLColor}

\newcommand{\beq}{\begin{equation}}
\newcommand{\eeq}{\end{equation}}
\newcommand{\ud}{\mathrm{d}}
\newcommand{\ui}{\mathrm{i}}
\newcommand{\ug}{\text{g}}
\newcommand{\um}{\text{m}}

\newcommand{\scP}{\mathscr{P}}

\newcommand{\RR}{\mathbb{R}}

\newcommand{\MM}{\mathcal{M}}
\newcommand{\veps}{\varepsilon}

\newcommand{\Lik}{\mathcal{L}_k}

\newcommand{\Lila}{\mathcal{L}_{\lambda}}
\newcommand{\Lixi}{\mathcal{L}_{\xi}}

\begin{document}

\title{First Law of Mechanics for Spinning Compact Binaries: \texorpdfstring{\\}{} Dipolar Order}

\author{Paul Ramond}\email{paul.ramond@obspm.fr}
\affiliation{Laboratoire Univers et Théories, Observatoire de Paris, CNRS, Université PSL, Université de Paris, F-92190 Meudon, France}

\author{Alexandre Le Tiec}\email{alexandre.letiec@obspm.fr}
\affiliation{Laboratoire Univers et Théories, Observatoire de Paris, CNRS, Université PSL, Université de Paris, F-92190 Meudon, France}

\date{\today}

\begin{abstract}
Building upon the Noether charge formalism of Iyer and Wald, we derive a variational formula for spacetimes admitting a Killing vector field, for a generic energy-momentum distribution with compact support. Applying this general result to the particular case of a binary system of spinning compact objects moving along an exactly circular orbit, modelled using the multipolar gravitational skeleton formalism, we derive a first law of compact binary mechanics at dipolar order. We prove the equivalence of this new result with the canonical Hamiltonian first law previously derived for binary systems of spinning compact objects, for spins colinear with the orbital angular momentum. This paper paves the way to an extension of the first law of binary mechanics to the next quadrupolar order, thereby accounting for the spin-induced and tidally-induced deformability of the compact bodies.
\end{abstract}

\maketitle

\section{Introduction}

\subsection{Motivation}

This paper is the second in a series of articles that aim at extending the first law of binary mechanics \cite{Fr.al.02,Le.al.12,Bl.al.13} to quadrupolar order, for binary systems of spinning compact objects with internal structure and moving along circular orbits. Mathematically, the hypothesis of an exactly closed circular orbit translates into the existence of an \textit{helical} Killing vector field $k^a$. In Ref.~\cite{RaLe.21} (hereafter Paper I), we introduced the multipolar gravitational skeleton formalism to model such spinning compact bodies, and derived a number of geometrical identities for such systems. In particular, we proved in Paper I that the helical Killing vector $k^a$ of the compact binary system is tangent to the worldline of each spinning particle, namely
\beq\label{k=zu}
    k^a|_\gamma = z u^a \, ,
\eeq
with $u^a$ the tangent 4-velocity to the worldline $\gamma$, normalized according to $u^a u_a = -1$, and $z$ the constant redshift parameter along $\gamma$. Moreover, it was proven in Paper I that for each spinning particle the 4-velocity $u^a$, the 4-momentum $p^a$ and the antisymmetric spin tensor $S^{ab}$ are all Lie-dragged along $k^a$, and thus along $\gamma$ thanks to the formula \eqref{k=zu}:
\beq\label{Lies}
    \Lik u^a = \Lik p^a = \Lik S^{ab} = 0 \, .
\eeq
In this paper, we will use those key results to derive a first law of binary mechanics at \textit{dipolar} order. The geometrical framework developed in this paper paves the way to an extension of the first law at the next quadrupolar order, thereby accounting for the rotationally-induced and tidally-induced deformability of the compact bodies.

Importantly, the first law of mechanics for spinning compact binaries \cite{Bl.al.13} (see also \cite{Fu.al.17,An.al.20}) has met various applications in the context of approximation methods such as post-Newtonian (PN) theory, the gravitational self-force (GSF) framework and the effective one-body (EOB) model \cite{Bl.14,BaPo.18}. Among those applications, the first law was used to:
\begin{itemize}
    \item[(i)] Compute the order-mass-ratio frequency shift of the innermost stable circular orbit of a particle orbiting a Kerr black hole induced by the conservative part of the GSF \cite{Is.al.14};
    \item[(ii)] Constraint various spin-dependent couplings that enter the effective Hamiltonian that controls the conservative dynamics of spinning compact-object binaries in a particular EOB model \cite{Bi.al.15};
    \item[(iii)] Validate analytical GSF calculations of the gyroscope precession of a spinning particle orbiting a Kerr black hole \cite{Bi.al.18} and of Detweiler's redshift parameter for a small extended compact body orbiting a Schwarzschild black hole \cite{Bi.al.20};
    \item[(iv)] Compare the predictions of the PN approximation and GSF theory to the numerical relativity results of sequences of quasi-equilibrium initial data for corotating black hole binaries \cite{LeGr.18};
    \item[(v)] Provide new spin-dependent contributions of the conservative dynamics for arbitrary-mass-ratio aligned-spin binary black holes at the fourth-and-a-half and fifth PN orders \cite{SiVi.20,An.al.20}.
\end{itemize}
See also \cite{BlLe.17} for a list of the wide range of applications of the first law of binary mechanics for nonspinning compact bodies, and Refs.~\cite{Po.al.20,Wa.al.21} for more recent works using the first law.

\subsection{Summary}

The first central result established in this paper is a general formula relating the variation $\delta H_\xi$ of the ``conserved quantity" canonically conjugate to the infinitesimal generator $\xi^a$ of an isometry, to the variations of two hypersurface integrals over the energy-momentum tensor $T^{ab}$ of all the compact-supported matter fields. This variational formula simply reads
\beq\label{bim!}
	\delta H_\xi = \delta \int_\Sigma \varepsilon_{abcd} \, T^{de} \xi_e - \frac{1}{2} \int_\Sigma \varepsilon_{abcd} \, \xi^d \, T^{ef} \delta g_{ef} \, ,
\eeq
where $\varepsilon_{abcd}$ is the canonical volume form associated with the metric $g_{ab}$, and $\Sigma$ is an arbitrary hypersurface transverse to $\xi^a$. Our main interest is in applying the general formula \eqref{bim!} to a binary system of compact objects moving on a circular orbit with constant angular velocity $\Omega$. The spacetime geometry is then invariant along the integral curves of a \textit{helical} Killing field of the form $k^a = t^a + \Omega \, \phi^a$, where $t^a$ is timelike and $\phi^a$ spacelike with closed integral curves of parameter length $2\pi$, and the variation of the associated ``conserved quantity" $H_k$ reads $\delta H_k = \delta M - \Omega \, \delta J$, where $M$ is the Arnowitt-Deser-Misner (ADM) mass and $J$ the total angular momentum of the binary system.

We use the gravitational skeleton framework to model each spinning compact body as a particle endowed with multipole moments defined along a timelike worldline $\gamma_\ui$ with unit tangent 4-velocity $u_\ui^a$, where the index $\ui \in \{1,2\}$ labels the two particles. Up to dipolar~order, each particle is entirely characterized by its 4-momentum $p^a_\ui$ and its (antisymmetric) spin tensor $S^{ab}_\ui$. The application of the general formula \eqref{bim!} then gives
\beq\label{yes!}
	\delta M - \Omega \, \delta J = \sum_\ui \left( z_\ui \, \delta m_\ui - \tfrac{1}{2} \, \nabla_a k^b \, \delta S^a_{\ui\,\, b} - k_a \delta \dot{D}_\ui^a + \tfrac{1}{2} \, \dot{D}_\ui^a \delta k_a \right) ,
\eeq
where for each particle we defined the rest mass $m \equiv - p^a u_a$, the redshift $z \equiv - k^a u_a$ and the mass dipole $D^a \!\equiv\! - S^{ab} u_b$, such that $D^a u_a \!=\! 0$. The overdot stands for the covariant derivative along the worldline $\gamma$, e.g. $\dot{D}^a \!\equiv\! u^c \nabla_c D^a$. The variational relation \eqref{yes!} is \textit{exact} to dipolar order in the multipolar gravitational skeleton framework; in particular, no perturbative expansion in powers of the spins has been performed. Interestingly, if the mass dipole $D^a$ is conserved, then the last two terms in the right-hand side of \eqref{yes!} are seen to vanish. This is guaranteed, for instance, by imposing the Frenkel-Mathisson-Pirani spin supplementary condition
\beq\label{Pirani}
	S^{ab} u_b = 0 \, .
\eeq

Alternatively, the first law of compact binary mechanics \eqref{yes!} can be expressed in terms of the conserved norms $|k|$, $|\nabla k|$ and $S_\ui$ of the helical Killing field $k^a$, the 2-form $\nabla_a k_b$ and the spin tensors $S_\ui^{ab}$. More precisely, by imposing the constraint \eqref{Pirani} and working to linear order in the spin amplitudes $S_\ui$, we find
\beq\label{ouhou!}
	\delta M - \Omega \, \delta J = \sum_\ui \bigl( |k| \, \delta m_\ui - |\nabla k| \, \delta S_\ui \bigr) + O(S^2) \, ,
\eeq
suggesting a pattern that might hold to all orders in a multipolar expansion, whereby higher-order multipolar contributions involve higher-order covariant derivatives of the helical Killing field. We then show that the form \eqref{ouhou!} of the first law is equivalent to the canonical Hamiltonian first law of binary mechanics derived in Ref.~\cite{Bl.al.13}, for canonical spins aligned or anti-aligned with the orbital angular momentum, to linear order in the spins.

The remainder of this paper is organized as follows. The general identity \eqref{bim!} is derived in Sec.~\ref{sec:varid}. The gravitational skeleton framework used to model each spinning particle, at dipolar order, is summarized in Sec.~\ref{sec:skeleton}. The main geometrical properties of helically symmetric spacetimes modelling binary systems of dipolar particles moving along circular orbits are discussed in Sec.~\ref{sec:helical}. The dipolar first law of binary mechanics \eqref{yes!} is then derived in Sec.~\ref{sec:firstlaw}. The following Sec.~\ref{sec:prec} is dedicated to a geometrical description of the spin precession of the spinning particles, which is used in Sec.~\ref{sec:Hamilton} to derive the form \eqref{ouhou!} of the first law, also shown to agree with the Hamiltonian first law of Ref.~\cite{Bl.al.13}. A summary of our conventions and notations, as well as a number of technical details, are relegated to appendices.

\section{Variational identity} \label{sec:varid}

In this section, we derive a general identity that relates the first-order variations of conserved asymptotic quantities in a diffeomorphism invariant theory of gravity---such as general relativity (GR)---to those of hypersurface integrals over the energy-momentum tensor of a generic distribution of matter with compact support. Following a short recollection of some preliminary results in  Sec.~\ref{subsec:preli}, we perform a gravity-matter split of the Lagrangian in Sec.~\ref{subsec:split}, out of which the variational identity is derived in Sec.~\ref{subsec:derivation}. The link to conserved asymptotic quantities is discussed in Sec.~\ref{subsec:asymptotic}, and the arbitrariness of the hypersurface of integration over the energy-momentum tensor is proven in Sec.~\ref{subsec:arbitrary}.

Throughout this section we shall use boldface symbols to denote differential forms defined over a 4-dimensional spacetime manifold. Given an arbitrary differential $p$-form $\bm{X}=X_{a_1\cdots a_p}$, its exterior derivative will be denoted $\ud \bm{X} = (\ud X)_{a_1\cdots a_{p+1}}$.

\subsection{Preliminaries}\label{subsec:preli}

Iyer and Wald \cite{Wa2.93,IyWa.94} gave a general derivation of the first law of black hole mechanics for arbitrary vacuum perturbations of a stationary black hole that are asymptotically flat at spatial infinity and regular on the event horizon. This derivation was extended to arbitrary electro-nonvacuum perturbations of charged black holes by Gao and Wald \cite{GaWa.01}, who further derived a ``physical process'' version of the first law. Here we follow their general strategy, while making appropriate modifications for our nonvacuum perturbations of a nonstationary spacetime with a generic, compact supported energy-momentum distribution. The following analysis will follow closely that of Iyer \cite{Iy.97}, except that our background spacetime will not be assumed to be a stationary-axisymmetric black hole solution. For simplicity and definiteness, we shall restrict our analysis to the classical theory of GR in four spacetime dimensions, but most of the calculations hold for a general diffeomorphism invariant theory of gravity in any dimension \cite{Iy.97}.

Our starting point is the Lagrangian of the theory, taken to be a diffeomorphism invariant 4-form $\bm{L}$ on spacetime, which depends on the metric $g_{ab}$ and other dynamical ``matter'' fields $\psi$, denoted collectively as $\phi \equiv (g,\psi)$. Let us consider a one-parameter family of spacetimes with metric $g_{ab}(\lambda)$. The first-order variation of $g_{ab}(\lambda)$ is defined as $\delta g_{ab} \equiv \ud g_{ab} / \ud \lambda|_{\lambda = 0}$, and similarly for other dynamical fields. The first-order variation of the Lagrangian $\bm{L}$ can always be written in the form \cite{LeWa.90,Wa2.93,IyWa.94,GaWa.01,Iy.97}
\beq\label{deltaL}
    \delta \bm{L} = \bm{E} \, \delta\phi + \ud \mathbf{\Theta}(\phi,\delta\phi) \, ,
\eeq
where summation over the dynamical fields (and contraction of the associated tensor indices) is understood in the first term on the right-hand side, and the Euler-Lagrange equations of motion can be read off as $\bm{E}(\phi) = 0$. The symplectic potential 3-form $\mathbf{\Theta}$ is a linear differential operator in the field variations $\delta\phi$. Because the Lagrangian is uniquely defined only up to an exact form, $\bm{L} \to \bm{L} + \ud \bm{\mu}$, the symplectic potential is defined only up to $\bm{\Theta} \to \bm{\Theta} + \delta \bm{\mu} + \ud \bm{Y}$, for some arbitrary 3-form $\bm{\mu}(\phi)$ and 2-form $\bm{Y}(\phi,\delta\phi)$.

Now, let $\xi^a$ denote an arbitrary smooth vector field on the unperturbed spacetime and $\Lixi$ the Lie derivative along $\xi^a$. From the Lagrangian $\bm{L}$ and its associated symplectic potential $\bm{\Theta}$, we define the Noether current 3-form $\bm{J}$ relative to $\xi^a$ according to
\beq\label{defJ}
    \bm{J}[\xi] \equiv \bm{\Theta}(\phi,\Lixi \phi)-\xi\cdot\bm{L} \, ,
\eeq
where $\bm{\Theta}(\phi,\Lixi \phi)$ stands for the expression of $\bm{\Theta}(\phi,\delta\phi)$ with each occurrence of $\delta \phi$ replaced by $\Lixi \phi$, and ``$\cdot$'' denotes the contraction of a vector field with the first index of a differential form, so that $\xi\cdot\bm{L} \equiv \xi^d L_{dabc}$. The key property of the Noether current \eqref{defJ} is that it is closed ($\ud\bm{J}=0$) if the field equations are satisfied ($\bm{E}=0$) or if the vector field $\xi^a$ Lie derives all of the dynamical fields ($\Lixi \phi = 0$). Indeed, taking the exterior derivative of Eq.~\eqref{defJ} readily gives \cite{LeWa.90,Ro.20}
\begin{align}\label{dJ}
    \ud\bm{J}[\xi] &= \ud \bm{\Theta}(\phi,\Lixi \phi)- \ud(\xi\cdot\bm{L}) \nonumber \\
    &= \Lixi \bm{L} - \bm{E} \, \Lixi \phi - (\Lixi \bm{L}-\xi\cdot \ud \bm{L}) \nonumber \\
    &=- \bm{E} \, \Lixi \phi \, ,
\end{align}
where in the second equality we used the Lagrangian variation $\Lixi\bm{L} = \bm{E} \, \Lixi \phi + \ud\bm{\Theta}(\phi,\Lixi\phi)$ [formally analogous to Eq.~\eqref{deltaL}] and Cartan's magic formula, as well as the identity $\ud \bm{L}=0$ in the third equality, since $\ud \bm{L}$ is a 5-form on a 4-dimensional manifold.

The form of the Noether current \eqref{defJ} can be further specified thanks to the identity \eqref{dJ}. Indeed, it can be shown \cite{IyWa.95,Iy.97} that there exists a 3-form (with an extra dual vector index) $\bm{C}_a(\phi)$ that is locally constructed out of the dynamical fields $\phi$ in a covariant manner, such that the rightmost term in \eqref{dJ} reads $\bm{E} \, \Lixi \phi = \ud (\bm{C}_a \xi^a)$, thus implying $\ud(\bm{J}[\xi] + \bm{C}_a \xi^a) = 0$. Consequently, according to the Poincar\'e lemma, there exists a 2-form $\bm{Q}[\xi]$ such that the Noether current \eqref{defJ} can locally be written in the form
\beq\label{J-dQ}
    \bm{J}[\xi] = - \bm{C}_a \xi^a + \ud \bm{Q}[\xi] \, .
\eeq
Crucially, $\bm{C}_a = 0$ whenever the field equations, $\bm{E} = 0$, are satisfied. One may view $\bm{C}_a = 0$ as being the constraint equations of the theory which are associated with its diffeomorphism invariance \cite{LeWa.90}. The ambiguity in $\bm{\Theta}$ discussed below \eqref{deltaL} implies that the Noether current is uniquely defined only up to $\bm{J}[\xi] \to \bm{J}[\xi] + \ud \bigl[ \bm{Y}(\phi,\Lixi\phi) \!-\! \xi \cdot \bm{\mu} \bigr]$ and the Noether charge up to $\bm{Q}[\xi] \to \bm{Q}[\xi] + \bm{Y}(\phi,\Lixi\phi) - \xi \cdot \bm{\mu}$. As shown in Ref.~\cite{IyWa.94}, those ambiguities will not affect the results stated in the following paragraphs, so from now on we shall omit them.

Next, we define the symplectic current 3-form by \cite{Wa2.93}
\beq\label{sc}
    \bm{\omega}(\phi,\delta_1 \phi, \delta_2 \phi) \equiv \delta_2 [\bm{\Theta}(\phi,\delta_1 \phi)] - \delta_1 [\bm{\Theta}(\phi,\delta_2 \phi)] \,,
\eeq
which depends on two linearly independent first-order variations $\delta_1\phi$ and $\delta_2\phi$ of the dynamical fields $\phi$. It can be shown that this differential form is closed ($\ud \bm{\omega} \!=\! 0$) when $\phi$ is a solution of the field equations and $\delta_1\phi$ and $\delta_2\phi$ are solutions of the linearized field equation \cite{LeWa.90,Ro.20}. The symplectic curent \eqref{sc} is used to define the notion of a Hamiltonian, which, in turn, gives rise to the notions of total energy and angular momentum \cite{LeWa.90,IyWa.94}.

Now, set $\delta_1 \phi \equiv \Lixi\phi$ and let $\delta_2 \phi \equiv \delta\phi$ correspond to a nearby solution for which $\delta \xi^a = 0$, as allowed by the diffeomorphism gauge freedom of GR. Then
\begin{align}\label{Omega}
    \bm{\omega}(\phi,\Lixi \phi,\delta \phi) &= \delta \bm{\Theta}(\phi,\Lixi \phi) - \Lixi \bm{\Theta}(\phi,\delta \phi) \nonumber \\
    &= \delta \bm{J}[\xi] + \xi \cdot \delta \bm{L} - (\xi \cdot \ud \bm{\Theta} + \ud (\xi \cdot \bm{\Theta})) \nonumber \\
    &= \ud \bigl( \delta \bm{Q}[\xi] - \xi \cdot \bm{\Theta} \bigr) + \xi \cdot \bm{E} \, \delta\phi - \delta (\bm{C}_a \xi^a) \, ,
\end{align}
where we used the definition \eqref{defJ} and Cartan's magic formula in the second equality, as well as the identities \eqref{deltaL} and \eqref{J-dQ} in the last equality. When the field equations are satisfied, $\bm{E} = 0$ and $\bm{C}_a = 0$ imply that the symplectic current \eqref{Omega} is exact and thus closed, as mentioned above. Integrating the resulting identity over a hypersurface $\Sigma$ transverse to $\xi^a$, with boundary $\partial{\Sigma}$, and using Stokes' theorem, we obtain the general formula \cite{Iy.97}
\beq\label{int}
	\int_\Sigma \bm{\omega}(\phi,\Lixi\phi,\delta\phi) = \int_{\partial{\Sigma}} \delta \bm{Q}[\xi] - \xi \cdot \bm{\Theta}(\phi,\delta\phi) \, .
\eeq
The symplectic current in Eq.~\eqref{Omega} is a linear differential operator in the field variation $\Lixi\phi$. Consequently, if $\xi^a$ Lie derives \textit{all} of the dynamical fields in the background ($\Lixi\phi = 0$), then the boundary integral on the right-hand side of Eq.~\eqref{int} vanishes identically.

\subsection{Gravity-matter split}\label{subsec:split}

To derive the general variational identity of interest, we shall further split the Lagrangian 4-form $\bm{L}$ of the theory into a purely gravitational (vacuum) part and a matter part:
\beq\label{L}
\bm{L}(g,\psi) \equiv \bm{L}_\ug(g)+\bm{L}_\um(g,\psi) \, .
\eeq
The vacuum GR Lagrangian $\bm{L}_\ug$ depends on the metric $g_{ab}$ and its derivatives and is explicitly given by $16\pi \bm{L}_\ug = R \, \bm{\varepsilon}$, where $R$ is the Ricci scalar and $\bm{\varepsilon}$ the canonical volume form associated with $g_{ab}$. The matter part $\bm{L}_\um$ is left unspecified, but is required to depend only on the metric $g_{ab}$ and the other dynamical ``matter'' fields $\psi$. Following Eq.~\eqref{deltaL}, the first-order variation of each Lagrangian in Eq.~\eqref{L} can then be split into a total derivative and a part related to the field equations, according to
\begin{subequations}\label{Lags}
    \begin{align}
        \delta \bm{L}_\ug &= - \frac{1}{16\pi} \, \bm{\varepsilon} \, G^{ab}\delta g_{ab} + \ud \mathbf{\Theta}_\ug(g,\delta g) \,, \label{Lagg} \\
        \delta \bm{L}_\um &= \frac{1}{2} \, \bm{\varepsilon} \, T^{ab} \delta g_{ab} + \bm{E}_\um(g,\psi) \, \delta \psi + \ud \mathbf{\Theta}_\um(\phi,\delta \phi) \, . \label{Lagsm}
    \end{align}
\end{subequations}
Here, $G_{ab} \equiv R_{ab} - \tfrac{1}{2} R g_{ab}$ is the Einstein tensor, $T_{ab}$ is the energy-momentum tensor and the matter field equations read $\bm{E}_\um(\phi) = 0$.

Repeating the analysis performed above in Sec.~\ref{subsec:preli}, separately for the (vacuum) gravity and matter sectors, one can easily show that the gravity and matter Noether currents take the form
\begin{subequations}
    \begin{align}
        \bm{J}_\ug[\xi] &\equiv \bm{\Theta}_\ug(g,\Lixi g) - \xi \cdot \bm{L}_\ug =  - \, \bm{C}^a_\ug \xi_a + \ud \bm{Q}_\ug[\xi] \, , \label{J-C-dQ_g} \\
        \bm{J}_\um[\xi] &\equiv \bm{\Theta}_\um(\phi,\Lixi\phi) - \xi \cdot \bm{L}_\um =  - \, \bm{C}_\um^a \xi_a + \ud \bm{Q}_\um[\xi] \, .
    \end{align}
\end{subequations}
On the one hand, the vacuum GR contributions are well known and explicitly read \cite{Wa2.93,IyWa.94}
\begin{subequations}
    \begin{align}
        \Theta^\ug_{abc}(g, \delta g) &= - \frac{1}{16\pi} \, \varepsilon_{abcd} \, g^{de} g^{fh} ( \nabla_f \delta g_{eh} - \nabla_e \delta g_{fh}) \, , \label{Theta_g} \\
        J^\ug_{abc}[\xi] &= - \frac{1}{8\pi} \, \varepsilon_{abcd} \nabla_e \nabla^{[e} \xi^{d]} \, , \label{J_g} \\
        Q^\ug_{ab}[\xi] &= - \frac{1}{16\pi} \, \varepsilon_{abcd} \nabla^c \xi^d \, , \label{Noether} \\
        C^\ug_{abce} &= - \frac{1}{8\pi} \, \varepsilon_{abcd} \, G^d_{\phantom{d}e} \, . \label{C_g}
    \end{align}
\end{subequations}
On the other hand, explicit forms of the matter contributions $\bm{\Theta}_\um$, $\bm{C}_a^\um$, $\bm{Q}_\um$ and $\bm{J}_\um$ depend on the particular choice of matter Lagrangian $\bm{L}_\um$. Typical examples include perfect fluids and electromagnetic fields \cite{Iy.97,Ro.20}.
Most importantly, whenever the matter field equations $\bm{E}_\um = 0$ are satisfied,
and regardless of the exact form of $\bm{L}_\um$, the matter constraint simply reduces to \cite{Iy.97}
\beq
    C^\um_{abce} = \varepsilon_{abcd} \, T^d_{\phantom{d}e} \, ,
\eeq
in such a way that the total constraint $\bm{C}_a \equiv \bm{C}_a^\ug + \bm{C}_a^\um$ in Eq.~\eqref{J-dQ} vanishes identically when the Einstein field equation $G_{ab} = 8\pi \, T_{ab}$ are satisfied as well.

\subsection{Variational identity}\label{subsec:derivation}

So far we considered an arbitrary smooth vector field $\xi^a$ defined on a background geometry $g_{ab}$. From now on, we shall further assume that $\xi^a$ is a \textit{Killing field} of the background, such that $\Lixi g_{ab} = 0$. Moreover, as allowed by the diffeomorphism gauge freedom of GR, we shall consider first-order variations for which
\beq\label{Gauge}
	\delta \xi^a = 0 \,,  \quad \text{implying} \quad \Lixi \delta g_{ab} = 0 \quad \text{and} \quad \Lixi \delta \xi_a = 0 \, .
\eeq
Consequently, the vacuum GR contribution $\bm{\omega}_\ug$ to the symplectic current \eqref{sc} can easily be shown to vanish identically. Indeed, from the explicit expression \eqref{Theta_g} for the pure gravity part $\bm{\Theta}_\ug$ of the symplectic potential 3-form, we have
\beq
    \bm{\omega}_\ug(g,\Lixi g,\delta g) = - \Lixi \bm{\Theta}_\ug(g,\delta g) = - \bm{\Theta}_\ug(g,\Lixi\delta g) = 0 \, ,
\eeq
where we used the Lie-dragging along $\xi^a$ of $g_{ab}$, $\delta g_{ab}$ and $\varepsilon_{abcd}$, as well as the commutation of the covariant derivative $\nabla_a$ and the Lie derivative $\Lixi$, as shown in Paper I. Consequently, only the matter part $\bm{\Theta}_\um$ contributes to \eqref{sc}, and we have
\begin{align}\label{Omega2}
    \bm{\omega}(\phi,\Lixi\phi,\delta\phi) &= \delta \bm{\Theta}_\um(\phi,\Lixi\phi) - \Lixi \bm{\Theta}_\um(\phi,\delta\phi) \nonumber \\
    &= \delta \bm{J}_\um[\xi] + \xi \cdot \delta \bm{L}_\um - (\xi \cdot \ud \bm{\Theta}_\um + \ud (\xi \cdot \bm{\Theta}_\um)) \nonumber \\
    &= \ud \bigl( \delta \bm{Q}_\um[\xi] - \xi \cdot \bm{\Theta}_\um \bigl) \;+\; \xi \cdot \bigl( \tfrac{1}{2} \, \bm{\varepsilon} \, T^{ab} \delta g_{ab} + \bm{E}_\um \, \delta\psi \bigr) - \delta (\bm{C}^a_\um \xi_a) \, ,
\end{align}
where we used the definition \eqref{defJ} and Cartan's magic formula in the second equality, as well as the Lagrangian variation \eqref{Lagsm} and \eqref{J-dQ} in the last equality. Equating the expressions \eqref{Omega} and \eqref{Omega2}, in which $\bm{Q}=\bm{Q}_\ug+\bm{Q}_\um$ and $\bm{\Theta}=\bm{\Theta}_\ug+\bm{\Theta}_\um$, and assuming the field equations are satisfied ($\bm{E} = 0$, $\bm{E}_\um = 0$ and $\bm{C} = 0$) implies the identity\footnote{If $\xi^a$ Lie derives all the dynamical fields in the background ($\Lixi \phi = 0$) and not merely the metric $g_{ab}$, then the right-hand side of \eqref{Omega} and \eqref{Omega2} vanish identically, yielding $\ud \bigl( \delta \bm{Q}_\ug[\xi] - \xi \cdot \bm{\Theta}_\ug \bigr) = - \ud \bigl( \delta \bm{Q}_\um[\xi] - \xi \cdot \bm{\Theta}_\um \bigr)$. This condition is stronger than merely equating the right-hand sides of Eqs.~\eqref{Omega} and \eqref{Omega2}, but it gives rise to the same identity \eqref{pilou-pilou}.}
\beq\label{pilou-pilou}
    \ud \bigl( \delta \bm{Q}_\ug[\xi] - \xi \cdot \bm{\Theta}_\ug \bigr) = \frac{1}{2} \, \xi \cdot \bm{\varepsilon} \, T^{ab} \delta g_{ab} - \delta (\bm{\varepsilon} \cdot T \cdot \xi) \, ,
\eeq
where we introduced the shorthand $\bm{\varepsilon} \cdot T \cdot \xi = \varepsilon_{abcd} T^{de} \xi_e$. Finally, integrating this equation over a hypersurface ${\Sigma}$ transverse to $\xi^a$, with boundary $\partial{\Sigma}$, and using Stokes' theorem, we obtain the simple identity
\beq\label{identity}
	\int_{\partial{\Sigma}} \delta \bm{Q}_\ug[\xi] - \xi \cdot \bm{\Theta}_\ug = \frac{1}{2} \int_{\Sigma} \xi \cdot \bm{\varepsilon} \; T^{ab} \delta g_{ab} - \delta \int_\Sigma \bm{\varepsilon} \cdot T \cdot \xi \, .
\eeq
This formula is very closely related to Eq.~(32) of Ref.~\cite{Iy.97}, valid for nonvacuum perturbations of stationary-axisymmetric black hole solutions in a general diffeomorphism invariant theory of gravity. Equation \eqref{identity} was also written down (without a detailed derivation) in Ref.~\cite{GrLe.13}, by adapting Refs.~\cite{IyWa.94,Iy.97,WaZo.00}, and applied to nonvacuum, nonstationary, nonaxisymmetric perturbations of a Kerr-black-hole-with-a-corotating-moon solution that is asymptotically flat at future null infinity.

\subsection{Asymptotic conserved quantities}\label{subsec:asymptotic}

Up to a numerical prefactor, the Noether charge $Q_\xi$ relative to the Killing vector field $\xi^a$ is defined as the integral of the Noether 2-form \eqref{Noether} over a topological 2-sphere $S$ that includes all the matter fields:
\beq\label{Komar_charge}
    Q_\xi \equiv \int_S \bm{Q}_\ug[\xi] \, .
\eeq
This charge is conserved in the sense that it does not depend on the choice of integration 2-surface $S$. Indeed, if $S$ and $S'$ denote two such topological 2-spheres, and $\Sigma$ any hypersurface bounded by $S$ and $S'$, then
\beq
    \int_S \bm{Q}_\ug[\xi] - \int_{S'} \bm{Q}_\ug[\xi] = \int_\Sigma \ud\bm{Q}_\ug[\xi] = - \frac{1}{8\pi} \int_\Sigma \varepsilon_{abcd} \nabla_e \nabla^{[e} \xi^{d]} \, = \frac{1}{8\pi} \int_\Sigma \varepsilon_{abcd}R^{de} \xi_e = 0 \, ,
\eeq
where we successively used Stokes' theorem, Eq.~\eqref{J-C-dQ_g} on shell with  \eqref{J_g}, the Kostant formula $\nabla_a \nabla_b \xi_c = R_{cbad} \xi^d$ (see Paper I), and the Einstein equation $R_{ab} = 0$ over the vacuum region $\Sigma$.

For an asymptotically flat spacetime with no isometry, the formula \eqref{Komar_charge} can be evaluated on a topological 2-sphere at spatial infinity. For instance, if $t^a$ and $\phi^a$ denote the \textit{asymptotic} Killing vectors associated with the invariance of an asymptotically Minkowskian spacetime under time translations and spatial rotations, then the Noether charge \eqref{Komar_charge} gives rise to the notions of Komar mass and Komar angular momentum
\beq\label{MK-JK}
	M_\text{K} \equiv 2 \int_\infty \bm{Q}_\ug[t] \quad \text{and} \quad J_\text{K} \equiv - \int_\infty \bm{Q}_\ug[\phi] \, .
\eeq
For an asymptotically flat spacetime, it can be established (see e.g. Ref.~\cite{Gou}) that the Komar angular momentum $J_K$ is equal to the ADM-like angular momentum $J$, also defined as a surface integral at spatial infinity, namely\footnote{In contrast to the ADM mass, energy and linear momentum, there is no such thing as “the ADM angular momentum.” One must impose additional asymptotic gauge conditions \cite{York.79} to have a unique, well-defined, ADM-type notion of angular momentum at spatial infinity; see e.g. Ref.~\cite{JaGo.11}.}
\beq\label{J_K=J}
    J_\text{K} = J \, .
\eeq
Under the additional assumption of stationarity, the equality $M_\text{K} = M$ of the Komar mass and the ADM mass was proven long ago \cite{Be.78,AsMa2.79}. This equality is closely related to a general relativistic generalization of the Newtonian virial theorem \cite{GoBo.94}, and was used as a criterion to compute quasi-equilibrium sequences of initial data for binary black holes \cite{Go.al.02,Gr.al.02,CoPf.04,An.05,Ca.al2.06,An.07}. Shibata \textit{et al.} \cite{Sh.al.04} showed that the equality $M_\text{K} = M$ holds for a much larger class of spacetimes; in particular, they could relax the restrictive hypothesis of stationarity.

When evaluated at infinity, the boundary term on the left-hand side of the identity \eqref{identity} has the natural interpretation of being the variation of the “conserved quantity” canonically conjugate to the asymptotic symmetry generated by $\xi^a$. Indeed, according to the analysis of Refs.~\cite{Wa2.93,IyWa.94,IyWa.95,WaZo.00} (see also \cite{Ro.20} for a review), if a Hamiltonian $H_\xi$ exists for the dynamics generated by the vector field $\xi^a$, then there exists a 3-form $\mathbf{B}_\ug$ such that
\beq\label{yep!}
	\int_\infty \delta \bm{Q}_\ug[\xi] - \xi \cdot \bm{\Theta}_\ug = \delta H_\xi \, , \quad \text{with} \quad H_\xi \equiv \int_\infty \bm{Q}_\ug[\xi] - \xi \cdot \mathbf{B}_\ug \, .
\eeq
Notably, for a stationary spacetime with a timelike Killing field $t^a$, normalized as $t^a t_a \to -1$ at spatial infinity, it can be shown that such a 3-form $\mathbf{B}_\ug$ exists, with $\int_\infty t \cdot \mathbf{B}_\ug = \tfrac{1}{2} M_\text{K} - M$, so that $H_t = M$. Similarly, for an axisymmetric spacetime with an axial Killing field $\phi^a$ one has $\int_\infty \phi \cdot \mathbf{B}_\ug = 0$, implying $H_\phi = - J$, in agreement with Eq.~\eqref{J_K=J}.

Finally, combining \eqref{identity} and \eqref{yep!} we conclude that if the hypersurface $\Sigma$ has no inner boundary (corresponding to the intersection of $\Sigma$ with a black hole horizon), then the variation of the conserved Noether charge associated with $\xi^a$ is related to the energy-momentum content through
\beq\label{variational_formula}
	\delta H_\xi = \delta \int_\Sigma \varepsilon_{abcd} \, T^{de} \xi_e - \frac{1}{2} \int_\Sigma \varepsilon_{abcd} \, \xi^d \, T^{ef} \delta g_{ef} \, .
\eeq
This variational formula is valid for a generic ``matter'' source with compact support. In this paper, we shall be interested in applying this general result to a binary system of spinning compact objects, modelled within the multipolar gravitational skeleton formalism reviewed in Paper I, up to dipolar order.

\subsection{Arbitrariness of the hypersurface} \label{subsec:arbitrary}

Before doing so, however, we show that the two integrals that appear in the right-hand side of \eqref{variational_formula} are independent of the choice of hypersurface $\Sigma$, and hence of ``time.'' Thereafter, it will prove convenient to introduce special notations for these hypersurface integrals, say
\begin{subequations}\label{integrals}
    \begin{align}
        I(\Sigma) &\equiv - \int_\Sigma \varepsilon_{abcd} \, T^{de} \xi_e = \int_\Sigma T^{ab} \xi_b \, \ud \Sigma_a \, , \label{I} \\
        K(\Sigma) &\equiv - \int_\Sigma \varepsilon_{abcd} \, \xi^d \, T^{ef} \delta g_{ef} = \int_\Sigma T^{ab} \delta g_{ab} \, \xi^c \ud \Sigma_c \, , \label{K}
    \end{align}
\end{subequations}
where $\ud \Sigma_a$ is the surface element normal to $\Sigma$. The integral $I$ is simply the flux across $\Sigma$ of the conserved Noether current $T^{ab} \xi_b$ associated with the Killing field $\xi^a$. The integral $K$, which involves the perturbed metric $\delta g_{ab}$, has no such simple physical interpretation.

Let $V$ denote a volume bounded by two spacelike hypersurfaces $\Sigma_1$ and $\Sigma_2$ and a worldtube that includes the support of $T^{ab}$. Then by using Stokes' theorem and the Leibniz rule, we readily find
\begin{subequations}\label{machin}
    \begin{align}
	    I(\Sigma_1) - I(\Sigma_2) &= \int_V \nabla_a (T^{ab} \xi_b) \, \ud V = \int_V \left[ (\nabla_a T^{ab}) \xi_b + T^{ab} \nabla_{(a} \xi_{b)} \right] \ud V = 0 \, , \label{diffI} \\
	    K(\Sigma_1) - K(\Sigma_2) &= \int_V \nabla_c (\xi^c T^{ab} \delta g_{ab}) \, \ud V = \int_V \left[ (\Lixi T^{ab}) \delta g_{ab} + T^{ab} \Lixi \delta g_{ab} \right] \ud V = 0 \, , \label{truc}
    \end{align}
\end{subequations}
where $\ud V$ is the invariant volume element. Here we used the local conservation of \mbox{energy and} momentum, $\nabla_a T^{ab} = 0$, together with Killing's equation $\nabla_{(a} \xi_{b)} = 0$, which implies $\nabla_c \xi^c = 0$, as well as $\Lixi T^{ab} = 0$ (see Paper I) and Eq.~\eqref{Gauge}.

Moreover, the integrals \eqref{integrals} are both invariant under Lie-dragging of the hypersurface $\Sigma$ in the direction of $\xi^a$. Given a spacelike hypersurface $\Sigma$ and a small positive number $\epsilon$, let $\Sigma_\epsilon$ denote the hypersurface obtained by Lie-dragging $\Sigma_0$ along the direction $\epsilon \xi^a$. With the shorthands $I_0 \equiv I(\Sigma_0)$, $I_\epsilon \equiv I(\Sigma_\epsilon)$ and $\dot{I} \equiv \lim_{\epsilon \to 0} \, (I_\epsilon - I_0)/\epsilon$, we have
\begin{subequations}
    \begin{align}
        \dot{I} &= - \int_{\Sigma_0} \Lixi (\varepsilon_{abcd} \, T^{de} \xi_e) = 0 \, , \\
        \dot{K} &= - \int_{\Sigma_0} \Lixi (\varepsilon_{abcd} \, \xi^d \, T^{ef} \delta g_{ef}) = 0 \, ,
    \end{align}
\end{subequations}
as a consequence of the Leibniz rule, $\Lixi \varepsilon_{abcd} = 0$ and $\Lixi T^{ab} = 0$ (see Paper I) and Eq.~\eqref{Gauge}. The fact that the right-hand side of Eq.~\eqref{variational_formula} is invariant under Lie-dragging in the direction of $\xi^a$ is consistent with the invariance \eqref{machin} of $I$ and $K$ on the choice of hypersurface.

\section{Dipolar particles}\label{sec:skeleton}

The multipolar gravitational skeleton formalism that is being used in this series of papers to model spinning compact objects was reviewed extensively in Paper I (see Sec.~II there). In this section, we shall first summarize briefly this model at dipolar order in Sec.~\ref{subsec:summary}, and then give more details on the consequences of our choice of spin supplementary condition in Sec.~\ref{subsec:SSC}.

\subsection{Energy-momentum tensor and equations of evolution}\label{subsec:summary}

At dipolar order, a spinning particle is entirely characterized by its 4-momentum $p^a$ and its antisymmetric spin tensor $S^{ab}$, which are both defined along a worldline $\gamma$ parameterized by the proper time $\tau$, with unit tangent 4-velocity $u^a$. The energy-momentum tensor of such a dipolar particle reads
\beq\label{SET}
    T^{ab} = \int_\gamma u^{(a} p^{b)} \, \delta_4 \, \ud \tau + \nabla_c \int_\gamma u^{(a} S^{b)c} \, \delta_4 \, \ud \tau \, ,
\eeq
where $\delta_4(x,x')$ is the invariant Dirac functional, a distributional biscalar defined so that for any smooth scalar field $f(x)$,
\beq\label{delta4}
    \int_V f(x) \, \delta_4(x,x') \, \ud V = f(x') \, ,
\eeq
where $V$ is a four-dimensional region of spacetime that contains the point $x'$; see App.~B in Paper I. For a given worldline $\gamma$, or equivalently for a given unit tangent $u^a$, the $4+6$ degrees of freedom contained in $(p^a,S^{ab})$ encode the same amount of information as $T^{ab}$ itself. The local conservation of energy and momentum, $\nabla_a T^{ab} \!=\! 0$, implies the Mathisson-Papapetrou-Dixon equations of evolution for $p^a$ and $S^{ab}$ along $\gamma$, which read
 \begin{subequations}\label{EE}
 		\begin{align}
		\dot{p}^a & = \frac{1}{2} R_{bcd}^{\phantom{dcb} a} S^{bc} u^d \, , \label{EoM} \\
 		\dot{S}^{ab} & = 2 p^{[a} u^{b]}  \, , \label{EoP}
 		\end{align}
 \end{subequations}
where the overdot stands for the covariant derivative along the direction $u^a$, e.g., $\dot{p}^a \!\equiv\! u^c \nabla_c p^a$. Notice that the spin coupling to curvature is orthogonal to the 4-velocity, such that $\dot{p}^a u_a = 0$. From the variables $u^a$, $p^a$ and $S^{ab}$, we introduce three scalar fields defined along $\gamma$: the rest mass $m$, the dynamical mass $\mu$ and the spin amplitude $S$, defined as
\begin{subequations}\label{defnorms}
    \begin{align}
        m &\equiv - p^a u_a \, , \label{m} \\
        \mu^2 &\equiv - p^a p_a \, , \label{mu} \\
        S^2 &\equiv \tfrac{1}{2} S^{ab} S_{ab} \, . \label{S}
    \end{align}
\end{subequations}
In general, the masses $m$ and $\mu$ need not coincide, as we shall see below. Finally, contracting Eq.~\eqref{EoP} with $u_a$ readily implies the following momentum-velocity relationship, which will prove useful in Sec.~\ref{sec:firstlaw} below to simplify the first law of binary mechanics:
	\beq \label{p=mu}
		p^a = m u^a - \dot{S}^{ab} u_b \, .
	\eeq

\subsection{Spin supplementary condition} \label{subsec:SSC}

The six degrees of freedom contained in the antisymmetric spin tensor $S^{ab}$ can equivalently be encoded in two spacelike vectors $S^a$ and $D^a$, both orthogonal to the 4-velocity $u^a$:
\beq\label{zip!}
	S^{ab} = \varepsilon^{abcd} u_c S_d + 2 D^{[a} u^{b]} \quad \Longleftrightarrow \quad
	\begin{cases}
		\, S^a \equiv - \frac{1}{2} \varepsilon^{abcd} u_b S_{cd} \, , \\
		D^a \equiv - S^{ab} u_b \, .
	\end{cases}
\eeq
Physically, the vector $D^a$ can be interpreted as the body's mass dipole moment, as measured by an observer with 4-velocity $u^a$, i.e., with respect to $\gamma$, while $S^a$ can be interpeted as the body's spin with respect to that wordline \cite{CoNa.15}. As explained in Paper I, in order to specify the worldline $\gamma$ uniquely, three constraints on the spin tensor $S^{ab}$, known as spin supplementary conditions (SSC), have to be imposed. In this paper, whenever we impose an SSC, we shall adopt the so-called Frenkel-Mathisson-Pirani SSC
\beq\label{SSC}
	D^a = 0 \quad \Longleftrightarrow \quad S^{ab} u_b = 0 \, .
\eeq
The first law of compact binary mechanics  will be shown in Sec.~\ref{sec:firstlaw} to take its simplest form whenever the SSC \eqref{SSC} holds.

\subsubsection{Conservation of spin amplitude and rest mass}

Together with the equations of evolution \eqref{EE}, the SSC \eqref{SSC} implies exact conservation laws. Indeed, contracting the equation of precession \eqref{EoP} with $S^{ab} = S^{[ab]}$ and using \eqref{SSC} shows that the particle's spin amplitude \eqref{S} is conserved:
\beq\label{dotS2}
	S \dot{S} = \tfrac{1}{2} S^{ab} \dot{S}_{ab} = S^{ab} p_{[a} u_{b]}  = - D^a p_a = 0 \, .
\eeq
Moreover, the particle's rest mass \eqref{m} is conserved along $\gamma$ as well. To prove this, we first apply the Leibniz rule to the rightmost term of Eq.~\eqref{p=mu} and use the SSC \eqref{SSC} to get
\beq \label{p=mu_bis}
	p^a = m u^a + S^{ab} \dot{u}_b \, .
\eeq
We stress that although Eq.~\eqref{p=mu} is valid without any SSC, Eq.~\eqref{p=mu_bis} holds only if $\dot{D}^a = 0$. Contracting \eqref{p=mu_bis} with $\dot{u}_a$ and using the orthogonality $u^a \dot{u}_a = 0$, as well as the antisymmetry of $S^{ab}$, yields $p^a \dot{u}_a = 0$. On the other hand, we already mentioned that $\dot{p}^a u_a = 0$. Therefore, we readily obtain the conservation along $\gamma$ of the rest mass $m$ defined in Eq.~\eqref{defnorms}:
	\beq \label{dotm}
		\dot{m} = - (p^a \dot{u}_a + \dot{p}^a u_a) = 0 \, .
	\eeq
Similar conservation laws hold for other choices of SSC. For instance, using the Tulczujew-Dixon SSC $S^{ab} p_b = 0$, one can establish the conservation of the spin amplitude $S$ and of the dynamical mass $\mu$ defined in \eqref{mu}. We emphasize that if the SSC $S^{ab} u_b = 0$ is imposed, the conservation laws \eqref{dotS2} and \eqref{dotm} are \textit{exact} at dipolar order, and not merely perturbatively valid in a power series expansion in the spin.

\subsubsection{Definition of a spin vector}

Having imposed the SSC \eqref{SSC}, the decomposition \eqref{zip!} implies that the three remaining degrees of freedom of the spin tensor $S^{ab}$ can be encoded in a spin vector $S^a$---or equivalently in a spin 1-form $S_a \equiv g_{ab} S^b$---obeying
\beq \label{spin_vector}
	S_a \equiv - \frac{1}{2} \varepsilon_{abcd} u^b S^{cd} \quad \Longleftrightarrow \quad S_{ab} = \varepsilon_{abcd} u^c S^d \, .
\eeq
By construction, the spin vector $S^a$ is spacelike and orthogonal to $u^a$, while its norm coincides with the conserved norm \eqref{defnorms} of the spin tensor, namely
\begin{subequations}
	\begin{align}
		& S^a u_a = 0 \, , \label{ortho_spinvec} \\
		& S^a S_a = S^2 \, . \label{norm_spinvec}
	\end{align}
\end{subequations}
The first result derives from the definition \eqref{spin_vector} and the antisymmetry of $\varepsilon_{abcd}$, while \eqref{norm_spinvec} derives from Eqs.~\eqref{spin_vector} and \eqref{defnorms} together with the SSC \eqref{SSC}. Moreover, the rate of change of the spin vector is easily computed from the definition \eqref{spin_vector} as
\beq\label{EoPvec}
	\dot{S}_a = - \frac{1}{2} \varepsilon_{abcd} \dot{u}^b S^{cd} \, ,
\eeq
where we used $\dot{\varepsilon}_{abcd} = 0$ by metric compatibility, the equation of spin precession \eqref{EoP}, and the antisymmetry of $\varepsilon_{abcd}$. This equation of evolution can be further simplified as follows. By substituting in \eqref{EoPvec} the expression \eqref{spin_vector} for $S^{cd}$ in terms of $u^a$ and $S^a$, and using the orthogonality $\dot{u}^a u_a = 0$, we readily obtain
	\beq \label{fermi-walker}
	\dot{S}^a = u^a \dot{u}^b S_b \, .
	\eeq
The spin vector $S^a$ is found to obey the Fermi-Walker transport law. This will be responsible for the Thomas precession that we shall encounter later in Sec.~\ref{subsec:liedragtetrad}.

\subsubsection{Momentum-velocity relations}

\begin{figure}[t!]
    \begin{center}
    	\includegraphics[width=0.2\linewidth]{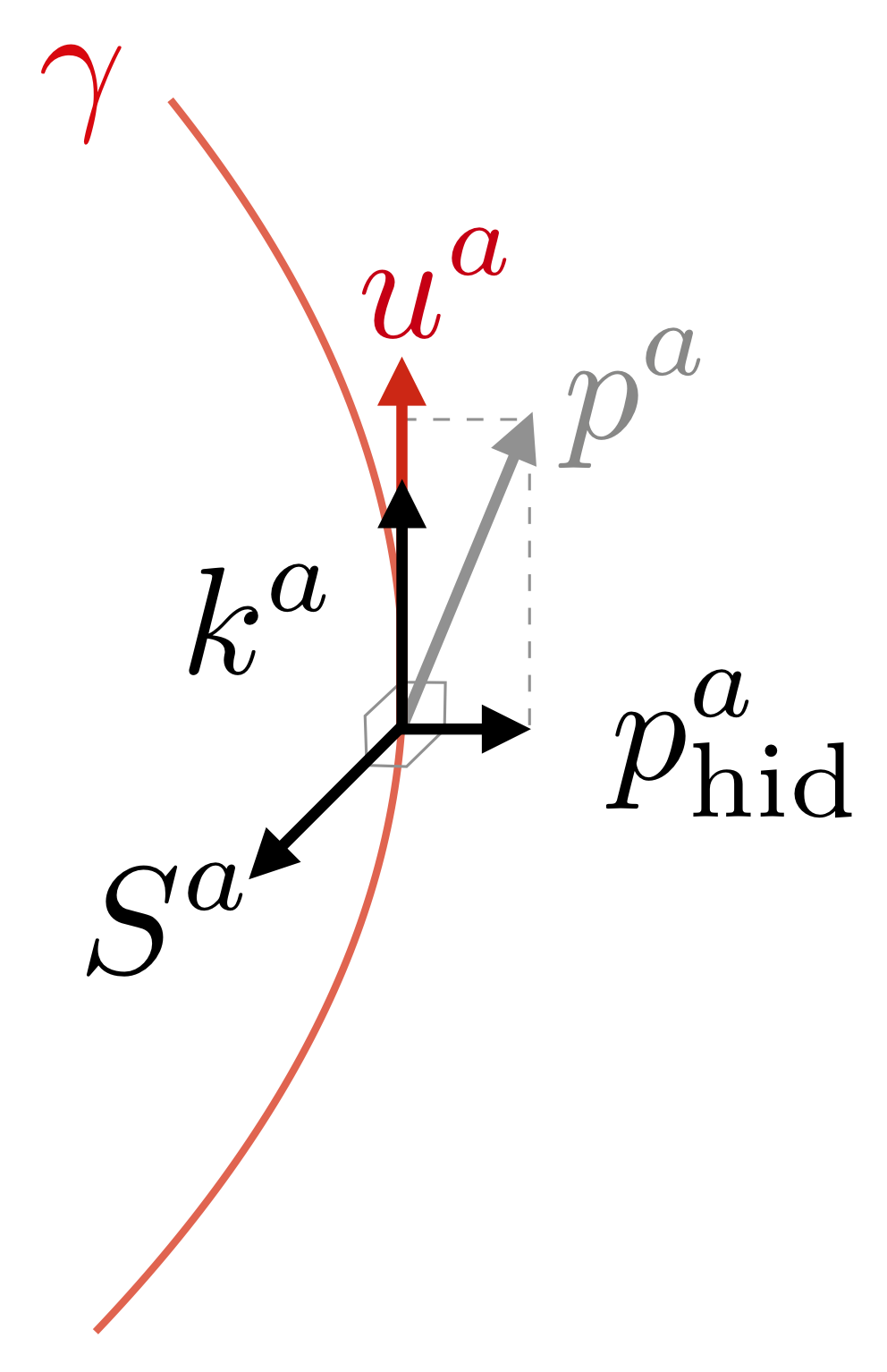}
        \caption{The worldline $\gamma$ of a spinning particle with tangent 4-velocity $u^a$, 4-momentum $p^a$, hidden momentum $p_\text{hid}^a$ and spin vector $S^a$. For a binary system of spinning particles moving on a circular orbit, the tangent 4-velocity of each particle is tangent to the helical Killing vector $k^a$ (see Sec.~\ref{sec:helical}).}
        \label{fig:particle}
    \end{center}
\end{figure}

The SSC \eqref{SSC} can be used to express the 4-velocity $u^a$ in terms of the 4-momentum $p^a$ and the antisymmetric spin tensor $S^{ab}$, thus closing the differential system \eqref{EE}, as expected. Indeed, the authors of Ref.~\cite{Co.al.18} recently established the momentum-velocity relation
\beq \label{mumvelo}
    m u^a = p^a + \frac{1}{S^2} S^{ab}S_{bc}p^c \,,
\eeq
where, interestingly, the rank-2 tensor $S^{ab}S_{bc}$ acts as a \textit{projector} orthogonal to both $u^a$ and $S^a$. Indeed, using Eq.~\eqref{spin_vector} and the identity $\varepsilon^{abcd} \varepsilon_{aefg} = - 6 \delta^{[b}_{\phantom{[b}e} \delta^{c}_{\phantom{c}f} \delta^{d]}_{\phantom{d]}g}$, (see, e.g., Ref.~\cite{Wal}), one has
\beq\label{projector}
    S^{ab}S_{bc} = - S^2 h^a_{\phantom{a}c} + S^a S_c \,,
\eeq
where $h^a_{\phantom{a}b} \equiv \delta^a_{\phantom{a}b} + u^a u_b$ is the projector orthogonal to the 4-velocity $u^a$. The relation \eqref{mumvelo} will not prove particularly useful for us. Instead, we derive an equivalent relation by substituting the expression \eqref{spin_vector} of the tensor $S^{ab}$ into the momentum-velocity relationship \eqref{p=mu_bis}. We find that the 4-momentum can alternatively be written as
\beq\label{p=mu_ter}
    p^a = m u^a - \varepsilon^a_{\phantom{a}bcd} u^b \dot{u}^c S^d \, ,
\eeq
which readily implies $p^a S_a = 0$. It is easily verified that the formulae \eqref{mumvelo} and \eqref{p=mu_ter} are equivalent \cite{Co.al.18}.\footnote{However, while the momentum-velocity relation \eqref{mumvelo} involves solely $u^a$, $p^a$ and $S^a$, the formulas \eqref{p=mu_bis} and \eqref{p=mu_ter} additionally involve the 4-acceleration $\dot{u}^a$.} From Eq.~\eqref{p=mu_ter}, the 4-momentum $p^a = p_\text{kin}^a + p_\text{hid}^a$ is the sum of the timelike kinematic momentum $p^a_\text{kin} \equiv m u^a$, with $m$ constant, and of the spacelike hidden momentum
\beq \label{phid}
    p^a_\text{hid} \equiv - \varepsilon^a_{\phantom{a}bcd} u^b \dot{u}^c S^d \,,
\eeq
which is orthogonal to $u^a$, $\dot{u}^a$ and $S^a$, as depicted in Fig.~\ref{fig:particle}. By using the condition of metric compatibility, which implies $\dot{\varepsilon}_{abcd} = 0$, as well as the equation of spin precession \eqref{fermi-walker}, the rate of change of the hidden momentum is simply given by
\beq \label{phidden}
    \dot{p}^a_\text{hid} \equiv - \varepsilon^a_{\phantom{a}bcd} u^b \ddot{u}^c S^d \, .
\eeq

\subsubsection{Acceleration and expansions in spin}

In contrast to the momentum-velocity relation \eqref{mumvelo}, the expression for the 4-acceleration $\dot{u}^a$ implied by the SSC \eqref{SSC} has been known for some time (see Ref.~\cite{KySe.07}) and reads \cite{Co.al.18}
\beq \label{accelnul}
\dot{u}^a = \frac{1}{S^2}\biggl( \frac{1}{m} B_{bc} S^b S^c S^a - S^{ab} p_b \biggr) \,,
\eeq
in which we introduced the gravito-magnetic part of the Riemann tensor, $B_{ab}$, defined as $B_{ab} \equiv \star R_{acbd} u^c u^d$, where $\star R_{abcd} \equiv \tfrac{1}{2} \varepsilon_{ab}^{\phantom{ab}ef} R_{efcd}$ is the self-dual of the Riemann tensor. (Notice that by using the definition \eqref{spin_vector} of the spin vector, the equation of motion \eqref{EoM} can be written in the more compact form $\dot{p}^a = B^{ab} S_b$.) Equation \eqref{accelnul} has a number of important consequences. First, it readily shows that $\dot{u}^a\neq 0$, implying that the motion of a spinning particle is not geodesic, in contrast to a test spin or a monopolar particle \cite{RamondPHD.21}. Second, \eqref{accelnul} is an exact formula for $\dot{u}^a$ in terms of $p^a$ and $S^{ab}$, provided that one uses Eq.~\eqref{mumvelo} to remove any dependence on $u^a$. Third, it gives a simple expression for the coefficient appearing in the Fermi-Walker transport law \eqref{fermi-walker} of the spin $S^a$, namely
\beq \label{FWv2}
    \dot{S}^a = \kappa u^a\,, \quad \text{with} \quad \kappa \equiv \frac{1}{m} B_{ab}S^aS^b \,,
\eeq
where we used $S^{ab}S_b=0$, which follows from the definition \eqref{spin_vector}. Therefore, the driving torque that prevents $S^a$ from being parallel-transported along $\gamma$ is quadratic in spin.

Next, the SSC \eqref{SSC} can be used to derive some approximate equations of evolution and algebraic relations, in the sense that they hold true only up to some order in the spin. This is particularly relevant since it is known that quadrupolar effects enter at the quadratic-in-spin level, and thus quadratic-in-spin effects in dipolar models are not self-consistent \textit{per se}. From now on, we shall denote by $O(S^n)$ any term that involves $n$ spin tensors (or spin vectors). To perform these expansions, instead of Eq.~\eqref{accelnul} we start by differentiating Eq.~\eqref{p=mu_ter} in the form $p^a=m u^a + p^a_{\text{hid}}$ and use Eqs.~\eqref{EoM} and \eqref{phidden} to find
\beq \label{dotu_0}
    \dot{u}^a = \frac{1}{m} B^{ab} S_b + \frac{1}{m} \varepsilon^a_{\phantom{a}bcd} u^b \ddot{u}^c S^d  = \frac{1}{m} B^{ab} S_b - \frac{1}{m} S^{ab} \ddot{u}_b \,,
\eeq
where the second equality simply follows from the formula \eqref{spin_vector}. Equation \eqref{dotu_0} is exact, and readily shows that $\dot{u}^a = O(S)$.
The rightmost term in Eq.~\eqref{dotu_0} is therefore at least of $O(S^2)$. Actually, the later involves $\ddot{u}_b$, and can therefore be expanded in powers of the spin at any order by recursively taking the covariant derivative along $u^a$ of Eq.~\eqref{dotu_0} and substituting it back into its own right-hand side. Doing this once, while using \eqref{FWv2} and the orthogonality $B^{ab} u_b = 0$, we can isolate the quadratic-in-spin contribution and obtain the following spin expansion:
\beq\label{dotu}
	\dot{u}^a = \frac{1}{m} B^{ab} S_b - \frac{1}{m^2} S^{ab} \dot{B}_{bc} S^c + O(S^3) \, .
\eeq
Equations \eqref{dotu_0} and \eqref{dotu} have a number of interesting consequences. First, if we substitute Eq.~\eqref{dotu} into the right-hand side of \eqref{p=mu_bis}, we obtain the following spin expansion for the 4-momentum:
	\beq \label{p=mu_quad}
		p^a = mu^a + \frac{1}{m} S^{ab} B_{bc} S^c + O(S^3) \, .
	\eeq
From this equation we see that $p^a = mu^a + O(S^2)$. Therefore the 4-momentum of a dipolar particle can only be aligned with its 4-velocity up to linear order in the spin. Second, taking the norm of Eq.~\eqref{p=mu_bis} provides a simple relation between the dynamical mass $\mu$ and the rest mass $m$: Using the SSC \eqref{SSC}, we readily obtain
\beq\label{mu-m}
	\mu^2 = m^2 - \dot{u}^a S_{ab} \dot{u}_c S^{cb}   \, .
\eeq
Equation \eqref{dotu} then implies $\mu^2 = m^2 + O(S^4)$, such that the two notions of mass coincide up to quartic-in-spin corrections. Since $\dot{m} = 0$ the dynamical mass satisfies $\dot{\mu} = O(S^4)$. Moreover, because the rightmost term in \eqref{mu-m} is the (squared) norm of the spacelike vector $\dot{u}_aS^{ab}$, it is positive and
\beq\label{mu2<m2}
    \mu^2 < m^2 \, .
\eeq
While this constraint does not necessarily imply $\mu^2 \!<\! 0$, the timelike nature of the momentum $p^a$ is expected, as pointed out in Ref.~\cite{Co.al.18}. Substituting for Eq.~\eqref{projector} into \eqref{mu-m} yields the alternative expression
\beq
    \mu^2 = m^2 - S^2 \dot{u}^a \dot{u}_a + (\dot{u}^a S_a)^2 \, ,
\eeq
which, combined with Eqs.~\eqref{FWv2} and \eqref{mu2<m2}, implies a \textit{lower} bound for the 4-acceleration for a given spin: $\dot{u}^2 > (\kappa/S)^2$. Finally, we emphasize one more time that the contributions of $O(S^2)$ in Eqs.~\eqref{dotu}--\eqref{p=mu_quad} are not self-consistent at dipolar order, because additional terms of $O(S^2)$ would contribute to those same equations if we were to include the spin-induced quadrupole in our gravitational skeleton model of spinning compact objects (see Paper I).

\section{Helical Killing symmetry}\label{sec:helical}

Our main interest is in applying the general variational formula \eqref{variational_formula} to the particular case of a binary system of spinning compact objects modelled as dipolar particles and moving along an exactly circular orbit. The approximation of a closed circular orbit translates into the existence of a helical Killing vector, which we define in Sec.~\ref{subsec:def}. We then discuss the issue of combining helical symmetry and asymptotic flatness in Sec.~\ref{subsec:asym_flat}, before exploring the general properties of helically symmetric binary systems of dipolar particles in Sec.~\ref{subsec:helical_binary}.

\subsection{Definition and properties}\label{subsec:def}

From now on, we shall consider spacetimes endowed with a global \textit{helical} Killing field $k^a$. Following the authors of Ref.~\cite{Fr.al.02} (see also \cite{Go.al.02}), a spacetime is said to have a helical Killing symmetry if the generator of the isometry can be written in the form
\beq\label{k}
	k^a = t^a + \Omega \, \phi^a \, ,
\eeq
where $\Omega > 0$ is a constant, $t^a$ is timelike and $\phi^a$ is spacelike with closed orbits of parameter length $2\pi$. In general, neither $t^a$ nor $\phi^a$ is a Killing vector, but the linear combination \eqref{k} is a Killing vector for a particular value of the constant $\Omega$, which can be interpreted as the angular velocity of the binary system.

\begin{figure}[t!]
    \begin{center}
    	\includegraphics[width=0.7\linewidth]{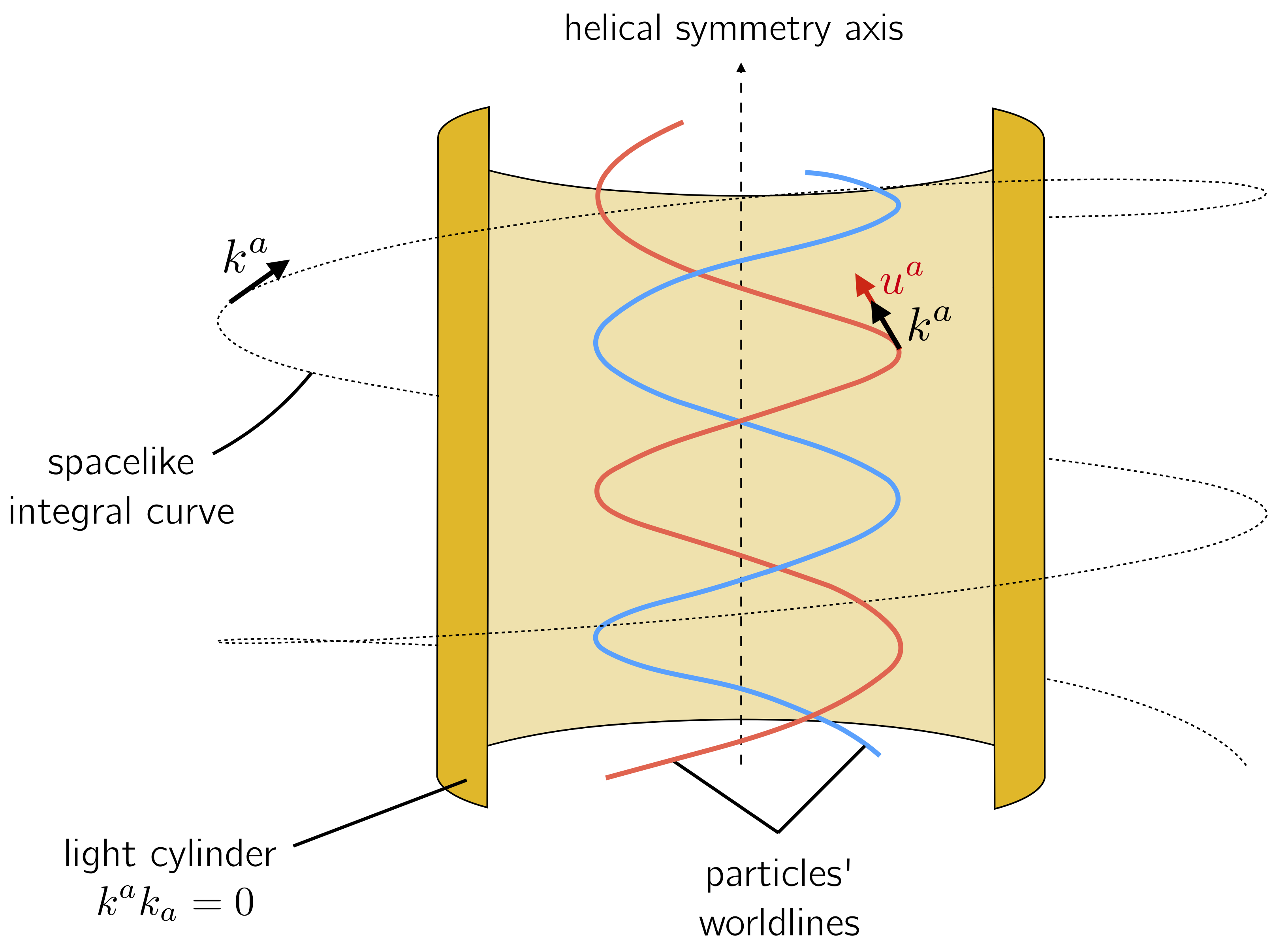}
        \caption{Main properties of a helical Killing field $k^a$. Integral curves that are spacelike lie outside the ``light-cylinder,'' where $k^a k_a = 0$. Within this cylinder, all integral curves are timelike and two of them represent the binary's motion on a circular orbit, as enforced by the condition $k^a|_\gamma \propto u^a$.}
        \label{fig:heli}
    \end{center}
\end{figure}

The null hypersurface over which $k^a k_a = 0$ is known as the light cylinder. Heuristically, it corresponds to the set of points where an observer would have a circular motion around the helical axis of symmetry ($\phi^a = 0$) with a ``velocity'' equal to the vacuum speed of light. Excluding any black hole region, the helical Killing field \eqref{k} is timelike everywhere inside the light cylinder and spacelike everywhere outside of it. These properties are summarized in Fig.~\ref{fig:heli}.

An important notion associated with the helical Killing vector field \eqref{k} is the \textit{twist}
\beq\label{twist}
    \varpi^a \equiv - \frac{1}{2} \varepsilon^{abcd} k_b \nabla_c k_d = 8\pi \, Q_\ug^{ab}[k] k_b \, ,
\eeq
where we recall that $Q^\ug_{ab}$ is the Noether charge 2-form associated with $k^a$, defined in \eqref{Noether}. The twist \eqref{twist} is orthogonal to $k^a$ and Lie-dragged along $k^a$, as a direct consequence of the Lie-dragging of $\varepsilon^{abcd}$, $k_b$ and $\nabla_c k_d$ (see Paper I). It will be shown in Sec.~\ref{subsec:prec_vort} that the twist \eqref{twist} does not vanish everywhere. As a consequence of the Frobenius theorem, $k^a$ cannot be  hypersurface-orthogonal. Moreover, it can be shown that the twist \eqref{twist} obeys the identity (see, e.g., Ref.~\cite{Wal})
\beq
    \nabla_{[a} \varpi_{b]} = \frac{1}{2} \varepsilon_{abcd} k^c R^{de} k_e \, .
\eeq
For a Ricci-flat, helically symmetric spacetime we thus have $\nabla_{[a} \varpi_{b]} = 0$, so that, locally, the 1-form $\varpi_a$ is exact, and there exists a scalar field $\varpi$ such that
\beq\label{omega=grad}
    \varpi_a = \nabla_a \varpi \,.
\eeq
The scalar twist $\varpi$ is then Lie-dragged along the helical Killing field $k^a$, as $\Lik \varpi = k^a \varpi_a = 0$.

For a binary system of dipolar particles moving along a circular orbit, the Ricci tensor vanishes everywhere, except along the worldlines of the particles, where the distributional energy-momentum tensor \eqref{SET} is singular. As explained in Paper I, the singular nature of a point-like source requires the introduction of a \textit{regularization method} (e.g. dimensional regularization \cite{HoVe.72,BoGi.72}) to remove the divergent self-field of each particle. In particular, any ``physically reasonable'' regularization method yields $T^{ab}|_\gamma = 0$ and $T|_\gamma = 0$, where $T \equiv g_{ab} T^{ab}$, so that $R^{ab}|_\gamma = 0$ by the Einstein field equation. Hence the spacetimes that we shall be considering in this paper are Ricci-flat in a regularized sense.\footnote{The helically symmetric, co-rotating binary black hole spacetimes that were considered in \cite{Fr.al.02,Gr.al.02,Go.al.02,LeGr.18} are Ricci-flat as well. The property \eqref{omega=grad} was not pointed out, however.}

\subsection{Helical isometry and asymptotic flatness}\label{subsec:asym_flat}

\hspace{-0.2cm}It has long been known that, in general relativity, helically symmetric spacetimes cannot be asymptotically flat \cite{Kl.04,GiSt.84,De.89}. This fact can easily be understood from a heuristic point of view: in order to maintain the binary on a fixed circular orbit, the energy radiated in gravitational waves needs to be compensated by an equal amount of incoming radiation. Far away from the source, the resulting system of standing waves ends up dominating the energy content of the spacetime, so that the falloff conditions necessary to ensure asymptotic flatness cannot be satisfied. As emphasized in Refs.~\cite{Fr.al.02,Sh.al.04,Le.al.12,GrLe.13,LeGr.18}, however, asymptotic flatness can be recovered if, loosely speaking, the gravitational radiation can be ``turned off.'' This can be achieved, in particular, using the Isenberg, Wilson and Mathews approximation to GR, also known as the conformal flatness condition (CFC) approximation \cite{IsNe.80,WiMa.89,Is.08}, or alternatively in the context of approximation methods such as PN theory \cite{Bl.14} and GSF theory \cite{BaPo.18}.

As written above, in general neither $t^a$ nor $\phi^a$ in \eqref{k} is a Killing vector. If the spacetime is asymptotically flat, however, then $t^a$ and $\phi^a$ are \textit{asymptotically} Killing, with the normalization $t^a t_a \to -1$ at infinity, and the surface integral at spatial infinity of the 2-form $\bm{Q}_\ug[k]$ yields the conserved charge associated with the generator \eqref{k}, namely [recalling Eq.~\eqref{MK-JK}]
\beq \label{kom1}
	\int_\infty \bm{Q}_\ug[k] = \frac{1}{2} M_\text{K} - \Omega J_\text{K} = \frac{1}{2} M_\text{K} - \Omega J  \, .
\eeq
The curious relative factor of two is related to the famous Komar ``anomalous factor'' entering the definitions of the Komar mass and angular momentum \cite{Ka.85,IyWa.94}.

Moreover, according to the general result \eqref{yep!}, the boundary term at spatial infinity in the identity \eqref{identity} yields a linear combination of the conserved charges associated with the asymptotic symmetry generators $t^a$ and $\phi^a$, namely\footnote{Asymptotically, the Killing vector field $\eqref{k}$ reduces to a linear combination of the generators $t^a$ and $\phi^a$ of time translations and spatial rotations, such that $\Omega$ should be treated as a constant while evaluating the surface integral \eqref{plouc}.}
\beq\label{plouc}
	\int_\infty \delta \bm{Q}_\ug[k] - k \cdot \bm{\Theta}_\ug = \delta \int_\infty \bm{Q}_\ug[k] - k \cdot \mathbf{B}_\ug = \delta H_k = \delta M - \Omega \, \delta J \, .
\eeq
The formula \eqref{plouc} is consistent with the results of Ref.~\cite{Fr.al.02}, obtained from a related analysis.

Finally, combining \eqref{identity} and \eqref{plouc}, we conclude that for a helically symmetric spacetime, the variations of the total mass and angular momentum are related to the energy-momentum content through
\beq\label{1st_law}
	\delta M - \Omega \, \delta J = \delta \int_\Sigma \varepsilon_{abcd} \, T^{de} k_e - \frac{1}{2} \int_\Sigma \varepsilon_{abcd} \, k^d \, T^{ef} \delta g_{ef} \, .
\eeq
The generalized first law \eqref{1st_law} is valid for a generic ``matter'' source with compact support, whose energy-momentum tensor $T^{ab}$ must be compatible with the helical isometry, $\Lik T^{ab} \!=\! 0$, as proven in Paper I. The variational formula \eqref{1st_law} holds, in particular, for perfect fluids. In App.~\ref{app:fluid} we show how \eqref{1st_law} reduces to the generalized first law derived by Friedman, Ury{\=u} and Shibata \cite{Fr.al.01}. In this paper, we shall be interested in applying this identity to a binary system of spinning compact objects, modelled within the multipolar gravitational skeleton formalism, up to dipolar order, as described in the previous Sec.~\ref{sec:skeleton}.

\subsection{Binary system of dipolar particles}\label{subsec:helical_binary}

In this subsection we consider a binary system of dipolar particles moving along a circular orbit, and explore some of the consequences of the Lie-dragging along the helical Killing vector \eqref{k} of the 4-velocity $u^a$, the 4-acceleration $\dot{u}^a$ and the spin vector $S^a$, as was established in Paper I. We first introduce the notion of vorticity and show that it is aligned with the spin vector. We then derive a simple closed-form formula for the 4-acceleration, which simplifies even further when the metric has a reflexion symmetry across an equatorial plane.

\subsubsection{Vorticity and spin vector}

First, for each particle we introduce the \textit{vorticity} $V^a$, namely the restriction to the worldline $\gamma$ of the twist \eqref{twist} associated with the helical Killing field \eqref{k}, defined as
\beq \label{vorticity}
	V^a \equiv - \frac{1}{2} \, \varepsilon^{abcd} u _b \nabla_c k_d \, .
\eeq
Indeed, the helical constraint \eqref{k=zu} implies $\varpi^a|_\gamma = z V^a$. The vorticity $V^a$ is orthogonal to $u^a$ and is Lie-dragged along $k^a|_\gamma \propto u^a$. The definition \eqref{vorticity} of the vorticity, which involves the Noether 2-form $\nabla_ak_b$, should be compared to the definition \eqref{spin_vector} of the spin vector, which involves the antisymmetric spin tensor $S_{ab}$. The duality between $(V^a,\nabla_ak_b)$ and $(S^a,S_{ab})$ is made even clearer when expressing $\nabla_a k_b$ in terms of $V^a$. Indeed, contracting Eq.~\eqref{vorticity} with $\varepsilon_{abcd}u^b$, using Eq.~\eqref{k=zu} and re-aranging the result readily gives
\beq \label{vort_invert}
	\nabla_a k_b|_\gamma = \varepsilon_{abcd} u^c V^d - 2 k_{[a} \dot{u}_{b]} \, ,
\eeq
which is to be compared to Eq.~\eqref{zip!}. We will use this formula in Sec.~\ref{sec:Hamilton} below to simplify the first law of compact binary mechanics.

We now consider the rate of change of the vorticity \eqref{vorticity} along $\gamma$. Using the condition of metric compatibility, which implies $\dot{\varepsilon}^{abcd} = 0$, together with the conservation of $\nabla_c k_d$ along $\gamma$ (see Paper I), we simply have $2\dot{V}^a = - \varepsilon^{abcd} \dot{u}_b \nabla_c k_d$. Substituting the decomposition \eqref{vort_invert} into this formula while using $\dot{u}^a u_a = 0$ then yields
\beq\label{Vdot}
	\dot{V}^a = u^a \dot{u}^b V_b \, ,
\eeq
showing that $V^a$ obeys the Fermi-Walker transport law, like the spin vector $S^a$. Contracting \eqref{Vdot} with $V_a$ then implies $\dot{V}^a V_a \!=\! 0$, so that the norm of the vorticity is conserved along $\gamma$.

The vorticity and the spin vector share one more property: provided that the SSC \eqref{SSC} is satisfied, they have the same spacelike direction. In order to establish this property, consider on the one hand the spacelike vector
\beq
    W^a \equiv \veps^a_{\phantom{a}bcd} u^b S^c V^d \, ,
\eeq
and on the other hand the Lie-dragging of the spin vector $S^a$ (see Paper I). The latter implies $z\dot{S}^a=S_b \nabla^b k^a$, an equation which can be simplified with the help of Eq.~\eqref{vort_invert}, to find an alternative expression for $W^a$ as
\beq \label{alt_W}
 W^a = z (\dot{S}^a - u^a \dot{u}^b S_b) \, ,
\eeq
which vanishes as a consequence of the equation of spin precession \eqref{fermi-walker}.\footnote{If the SSC \eqref{SSC} is not imposed, then the relationship $S^{[a} V^{b]} = 0$ can be generalized to $S^{[a} V^{b]} = z D^{[a} \dot{u}^{b]}$, as can easily be shown by generalizing the equation of spin precession \eqref{fermi-walker} to a nonzero mass dipole $D^a$.} Since $S^a$ and $V^a$ are both orthogonal to $u^a$, the equation $W^a = 0$ holds if, and only if, $S^a$ and $V^a$ are aligned. Let $s^a$ denote their common unit spacelike direction, such that $s^a s_a = 1$. Then we obtained the important result
\begin{subequations}\label{V-S}
	\begin{align}
		V^a &= V s^a \, , \\
		S^a &= S \, s^a \, ,
	\end{align}
\end{subequations}
where $V \equiv (V^a V_a)^{1/2}$ is the norm of the vorticity and $S$ that of the spin vector, as defined in Eq.~\eqref{norm_spinvec}. The norms $V$ and $S$ are both constant along $\gamma$, while $s^a$ is Lie-dragged along $\gamma$. The colinearity \eqref{V-S} will prove useful in Sec.~\ref{sec:firstlaw} to write the first law of binary mechanics in its simplest form, in terms of scalar quantities.

\subsubsection{Four-acceleration}

For a spinning particle in a binary system on a circular orbit, we may use the Lie-dragging of the 4-acceleration $\dot{u}^a$ along $k^a$, namely $z\ddot{u}_b = \dot{u}^c \nabla_c k_b$ (see Paper I), to express Eq.~\eqref{dotu_0} in the implicit form
\beq\label{cool}
    m \dot{u}^a = (P^{-1})^a_{\phantom{a}b} B^{bc} S_c \, , \quad \text{where} \quad P^a_{\phantom{a}b} = \delta^a_{\phantom{a}b} + \frac{1}{zm} \, S^{ac} \nabla_b k_c \, .
\eeq
The term linear in the spin tensor in the operator $P^a_{\phantom{a}b}$ can be rewritten in terms of the spin vector $S^a$ and the vorticity $V^a$ by substituting the expressions \eqref{spin_vector} and \eqref{vort_invert} for $S^{ab}$ and $\nabla_a k_b$. Using the colinearity \eqref{V-S} and the projector $h^a_{\phantom{a}b} \equiv \delta^a_{\phantom{a}b} + u^a u_b$ orthogonal to the 4-velocity, we readily find
\beq\label{SDk}
    S^{ac} \nabla_b k_c = p_\text{hid}^a k_b + VS \, (h^a_{\phantom{a}b} - s^a s_b) \, ,
\eeq
where we recall that $p_\text{hid}^a = - \varepsilon^a_{\phantom{a}bcd} u^b \dot{u}^c S^d$ is the spacelike ``hidden momentum" appearing in the momentum-velocity relationship \eqref{p=mu_ter}, and $h^a_{\phantom{a}b} - s^a s_b$ is the projector in the spacelike plane orthogonal to the common axial direction $s^a$ of $S^a$ and $V^a$. If $e_1^a$ and $e_2^a$ denote two spacelike unit vectors spanning that plane, such that $(u^a,e_1^a,e_2^a,s^a)$ is an orthonormal tetrad (cf. Sec.~\ref{subsec:orthotetrad} below), then the operator $P^a_{\phantom{a}b}$ in Eq.~\eqref{cool} reads
\beq\label{P}
    P^a_{\phantom{a}b} = \delta^a_{\phantom{a}b} + \frac{1}{m} p_\text{hid}^a u_b + \frac{VS}{zm} \, (e_1^a e_{1b} + e_2^a e_{2b}) \, . \,
\eeq

Remarkably, the inverse $(P^{-1})^a_{\phantom{a}b}$ of the operator \eqref{P} can be written in closed form by assuming an Ansatz of the form $(P^{-1})^a_{\phantom{a}b} = \delta^a_{\phantom{a}b} + \alpha \, p_\text{hid}^a u_b +\beta \, (e_1^a e_{1b} + e_2^a e_{2b})$, with $\alpha,\beta$ two constants to be solved for. Thanks to the defining identity $(P^{-1})^a_{\phantom{a}b} P^b_{\phantom{b}c} =\delta^a_{\phantom{c}c}$, the orthonormality of the diad $(e_1^a,e_2^a)$, and the orthogonality properties $p^a_{\text{hid}}u_a=0=p^a_{\text{hid}}s_a$, one obtains
\beq\label{P^-1}
    (P^{-1})^a_{\phantom{a}b} = \delta^a_{\phantom{a}b} - \frac{1}{1+a\omega} \, \frac{1}{m} p_\text{hid}^a u_b - \frac{a\omega}{1 + a\omega} \, (e_1^a e_{1b} + e_2^a e_{2b}) \, .
\eeq
Here, we introduced the Kerr parameter $a \equiv S/m$ of the spinning particle and we anticipated on the key formula $V = z \omega$, with $\omega$ the invariant spin precession frequency to be defined in Sec.~\ref{sec:prec} below; see e.g. Eqs.~\eqref{normomega}, \eqref{covomega} and \eqref{V=zomega}. Substituting Eq.~\eqref{P^-1} back into the expression \eqref{cool} for the 4-acceleration while using $u_b B^{bc} = 0$ then gives
\beq\label{udot_Killing}
    \dot{u}^a = \frac{a}{1+a\omega} \left( B^{ab} s_b + a\omega s^a B^{bc} s_b s_c \right) .
\eeq
This simple formula shows that the 4-acceleration is entirely sourced by the coupling of the magnetic-type tidal field with the spin vector, just like the rate of change $\dot{p}^a = B^{ab} S_b$ of the 4-momentum. Equations \eqref{ortho_spinvec} and \eqref{udot_Killing} yield $\dot{S}^a u_a = - \dot{u}^a S_a = - \kappa$, in agreement with the spin precession equation \eqref{FWv2}. Using the orthonormal triad $(e_1^a,e_2^a,s^a)$, the 4-acceleration \eqref{udot_Killing} can be expanded according to
\beq\label{plac}
    \dot{u}^a = \frac{a}{1+a\omega} \left( B_{13} \, e^a_1 + B_{23} \, e^a_2 \right) + a B_{33} \, s^a \, ,
\eeq
where $B_{IJ} = B_{ab} e_I^a e_J^b$ are the triad components of the gravito-magnetic part of the Riemann tensor. Provided that the legs $e_1^a$ and $e_2^a$ are Lie-dragged along $\gamma$ (see Sec.~\ref{subsec:liedragtetrad} below), those triad components are conserved, i.e. $\dot{B}_{IJ} = 0$, as a consequence of the Lie-dragging of $\varepsilon_{abcd}$, $R_{abcd}$, $u^a$ and $S^a$ (see Paper I).

Finally, we can substitute Eq.~\eqref{udot_Killing} into \eqref{p=mu_ter} to obtain a momentum-velocity relation of the form $p=f(u,S)$, namely
\beq
    p^a = m u^a + \frac{\varepsilon^{abcd} u_b S_c B_{de} S^e}{m + S \omega} \, ,
\eeq
which can be seen as the ``inverse" of Eq.~\eqref{mumvelo}, in the specific case of a helically-symmetric spacetime for a binary system of spinning particles.

\subsubsection{Reflection symmetry across the equatorial plane}\label{subsubsec:reflexion}

Under some circumstances, which will be detailed in Sec.~\ref{subsec:aligned} below, the Lie-dragging of the spin vector implies that the metric has a (discrete) reflection symmetry with respect to an equatorial plane, which coincides with the orbital plane of the binary system of spinning particles. The analysis of Ref.~\cite{Do.al.15} then implies the constraint
\beq\label{BSS}
    B_{33} = B_{ab} s^a s^b = 0 \, ,
\eeq
which has two important consequences. First, the spin evolution equation \eqref{FWv2} reduces to $\dot{S}^a = 0$, so that the spin is parallel-transported along $\gamma$, in addition to being Lie-dragged along $\gamma$. Second, the formula \eqref{udot_Killing} simplifies even further, and shows that the 4-acceleration belongs to the spacelike 2-space orthogonal to $u^a$ and $S^a$, according to
\beq\label{udot_Killing_bis}
    \left( m + S\omega \right) \dot{u}^a = B^{ab} S_b \, .
\eeq
The formula \eqref{udot_Killing_bis} is remarkably analogous to the equation of motion $\dot{p}^a \!=\! B^{ab} S_b$ for spinning particles, up to a ``renormalization" of the mass, $m \to m + S\omega$. Together with $\dot{S}^a = 0$, the formula \eqref{udot_Killing_bis} readily implies $(m + S\omega) \, \ddot{u}^a = \dot{B}^{ab} S_b$.

\section{First law at dipolar order} \label{sec:firstlaw}

In this section we derive the first law of mechanics, for a binary system of dipolar particles moving along an exactly circular orbit. We begin in Sec.~\ref{subsec:hypersurface} by using the freedom to choose the spacelike hypersurface of integration to simplify the calculations. Then, from the integral form of the first law derived in Sec.~\ref{sec:varid}, we compute in Sec.~\ref{subsec:scalar} the two integrals that appear in the right-hand side of \eqref{1st_law}. The resulting expressions are simplified algebraically in Sec.~\ref{subsec:algebraic}, and those results are combined in Sec.~\ref{subsec:combine} to establish the final formula. Throughout this section we do \textit{not} impose the SSC \eqref{SSC} to keep the results as general as possible.

\subsection{Choice of spacelike hypersurface}\label{subsec:hypersurface}

We proved in Sec.~\ref{subsec:arbitrary} that the right-hand side of the identity \eqref{1st_law} does not depend on the choice of spacelike hypersurface $\Sigma$. We may thus conveniently choose this hypersurface such that, for each particle, the Killing field $k^a$ is orthogonal to $\Sigma$ at the intersection point $\scP \equiv \Sigma \cap \gamma$, i.e.
\beq\label{k=|k|n}
	k^a \stackrel{\scP}{=} |k| n^a \, ,
\eeq
where $n^a$ is the future-directed, unit normal to ${\Sigma}$. Since $k^a|_\gamma = z u^a$, this implies $n^a|_\scP = u^a$ and $|k|_\scP = z$, where the redshift parameter $z$ was shown in Paper I to be constant along $\gamma$. More generally, one may introduce a foliation of the spacetime manifold $\mathcal{M}$ by a family of spacelike hypersurfaces $(\Sigma_t)_{t \in \mathbb{R}}$ such that Eq.~\eqref{k=|k|n} holds at \textit{any} point $\scP \in \gamma$, so that
\beq\label{k=|k|n_bis}
	k^a \stackrel{\gamma}{=} |k| n^a \, ,
\eeq
where $n^a$ is the future-directed, unit normal to the family $(\Sigma_t)_{t \in \mathbb{R}}$ of spacelike hypersurfaces, such that $n^a|_\gamma = u^a$. Along $\gamma$, the norm $|k|_\gamma = z$ of the helical Killing field then plays the role of a constant lapse function, and $k^a$ that of the normal evolution vector (see, e.g., \cite{Go.07}). According to Paper I, we then have $\Lik n^a|_\gamma = 0$, such that the extrinsic curvature vanishes at any point along $\gamma$:
\beq\label{Kab}
    K_{ab} \equiv - \frac{1}{2} \mathcal{L}_n \gamma_{ab} \stackrel{\gamma}{=} - \frac{1}{2|k|} \Lik \gamma_{ab} \stackrel{\gamma}{=} - \frac{1}{|k|} n_{(a} \Lik n_{b)} \stackrel{\gamma}{=} 0 \, ,
\eeq
where $\gamma_{ab} \equiv g_{ab} + n_a n_b$ is the induced metric on any spacelike hypersurface of the foliation, so that $n^a \gamma_{ab} = 0$. By combining \eqref{Kab} with the equalities $n_a|_\gamma = u_a$ and $n'_a|_\gamma \equiv n^c \nabla_c n_a|_\gamma = \dot{u}_a$, the usual 3+1 formula for the gradient of the unit normal to $\Sigma$ becomes \cite{Go.07}
\beq\label{gradient}
    \nabla_a n_b = - K_{ab} - n_a n'_b \stackrel{\gamma}{=} - u_a \dot{u}_b \, .
\eeq
We emphasize that Eqs.~\eqref{k=|k|n_bis}--\eqref{gradient} are valid only along the worldline $\gamma$, and thus in particular at the intersection point $\scP = \Sigma \cap \gamma$ with a given hypersurface $\Sigma$. Most importantly, one cannot choose the hypersurface $\Sigma$ such that \eqref{k=|k|n} holds in an open neighborhood of $\scP$. As explained in Sec.~\ref{subsec:prec_vort}, this is closely related to the helical nature of the Killing vector field \eqref{k}, whose twist \eqref{twist} does not vanish everywhere, which by Frobenius' theorem cannot be hypersurface orthogonal.

\subsection{Hypersurface integrals}\label{subsec:scalar}

In this subsection, we provide integrated expressions for the integrals $I$ and $K$ that appear in the right-hand side of the first law, as given in the form \eqref{1st_law}. To avoid being repetitive, we shall detail the calculation for $I$ only, and merely quote the final result for $K$.

We begin by substituting the dipolar energy-momentum skeleton \eqref{SET} into the definition \eqref{I} of $I$ to obtain
\beq\label{doulbint}
	I = \int_{\Sigma} \int_{\gamma} u^{(a} p^{b)} k_b \, \delta_4 \, \ud \tau \ud \Sigma_a + \int_{\Sigma} \int_{\gamma} \nabla_c \bigl( u^{(a} S^{b)c} \delta_4 \bigr) k_b  \, \ud \tau \ud \Sigma_a \, .
\eeq
To evaluate those two integrals, it is convenient to consider the foliation $(\Sigma_t)_{t\in\RR}$ introduced above, and to choose for $\Sigma$ one of the leafs of this foliation, say $\Sigma_{t_0}$ for some fixed $t_0\in\RR$. We recall that $n^a$ denotes the future-directed, unit normal to any leaf $\Sigma_t$ and $\gamma_{ab} = g_{ab} + n_a n_b$ the induced metric on $\Sigma_t$. After performing the change of variable $\tau\rightarrow t$, while using the standard 3+1 formulas (see e.g. Ref.~\cite{Go.07}) $\ud \tau = N \ud t$, $\ud \Sigma_a = - n_a \sqrt{\overline{\gamma}} \, \ud^3 x$ and $N\sqrt{\overline{\gamma}}=\sqrt{-g}$, where $N$ is the lapse function, $\overline{\gamma} \equiv \text{det}\,\gamma_{ij}$ and $g \equiv \text{det}\,g_{\alpha\beta}$, we obtain
\beq\label{KI_V}
	I = - \int_{\mathcal{M}} u^{(a} p^{b)} k_b n_a \, \delta_4 \, \ud V - \int_{\mathcal{M}} \nabla_c \bigl( u^{(a} S^{b)c} \delta_4 \bigr) k_b  n_a \, \ud V  \, ,
\eeq
where $\ud V = \sqrt{-g} \, \ud t \, \ud^3 x$ is the invariant 4-volume element over $\mathcal{M} = \bigcup_{t\in\RR} \Sigma_t$. The first integral in \eqref{KI_V} can readily be evaluated by using the defining property \eqref{delta4} of the invariant Dirac distribution $\delta_4$. Using the Leibniz rule and applying Stokes' theorem to the second integral yields
\beq\label{piloupilou}
    \int_{\MM} \nabla_c \bigl( u^{(a} S^{b)c} \delta_4 \bigr) k_b  n_a \, \ud V = \int_{\partial\MM} u^{(a} S^{b)c} k_b  n_a \, \delta_4 \, \ud \Sigma_c - \int_{\MM} u^{(a} S^{b)c} \nabla_c(k_b  n_a) \,  \delta_4 \, \ud V \, .
\eeq
The boundary term vanishes because its support is restricted to the single point $\scP = \gamma \cap \Sigma$, which does not intersect the boundary $\partial\MM$ of the manifold $\MM$. The remaining integral over $\MM$ in \eqref{piloupilou} can be evaluated once again by means of the property \eqref{delta4}. Finally, we obtain the expression
\beq\label{K_1}
	I = - u^{(a} p^{b)} k_b n_a + u^{(a} S^{b)c} \nabla_c (k_b  n_a) \, .
\eeq

For the integral $K$ defined in Eq.~\eqref{K} we follow the exact same steps, i.e., we perform a 3+1 decomposition, integrate by parts, apply Stokes' theorem, and lastly we use Eq.~\eqref{delta4}. We obtain the integrated formula
\beq\label{I_1}
	K = p^a k^b \delta g_{ab} + u^b S^{cd} \nabla_d(k^a  n_a \delta g_{bc}) \, .
\eeq
It should be understood that Eqs.~\eqref{K_1}--\eqref{I_1} are to be evaluated at the point $\scP$. Therefore, in the first term in the right-hand side of Eq.~\eqref{K_1}, one may freely use \eqref{k=|k|n}, which implies in particular $n^a|_\scP = u^a$. However, those relations cannot be used in the second terms in the right-hand sides of Eqs.~\eqref{K_1} and \eqref{I_1}, because the formula \eqref{k=|k|n} is only valid at $\scP$, and not in an open neighborhood of $\scP$; recall the remark below Eq.~\eqref{gradient}. Finally, we note that the expressions \eqref{K_1} and \eqref{I_1} hold irrespective of a particular choice of SSC.

\subsection{\texorpdfstring{Algebraic reduction of $I$ and $K$}{}}\label{subsec:algebraic}

We shall now simplify algebraically the expressions \eqref{K_1}--\eqref{I_1} for the integrals $I$ and $K$. We start with the result \eqref{K_1}. First, as $u^a n_a =-1$ and $k_b n_a = k_a n_b$ at $\scP$ by virtue of \eqref{k=|k|n}, the first term is simply $p^a k_a$. For the second term, we expand the symmetry in $u^{(a} S^{b)c}$ and the gradient  $\nabla_c(k_b n_a)$ by the Leibniz rule. This gives
\beq\label{K_2}
    I = p^a k_a + \frac{1}{2} ( u^a S^{bc} + u^b S^{ac} ) ( n_a \nabla_c k_b + k_b \nabla_c n_a ) \, .
\eeq
By substituting the formula \eqref{gradient} into Eq.~\eqref{K_2}, while using Killing's equation $\nabla_{(a} k_{b)} = 0$, the helical constraint $k^a|_\gamma = zu^a$, which implies $u^a k_a \!=\! -z$ and $\dot{k}^a = zu^a$, as well as $n^a|_\scP = u^a$ and the orthogonality $u^a \dot{u}_a = 0$, we readily obtain the simple expression
\beq\label{K_3}
    I = p^a k_a - D^a \dot{k}_a + \frac{1}{2} S^{ab} \nabla_a k_b  \, ,
\eeq
where we recall that $D^a = - S^{ab} u_b$ is the mass dipole moment with respect to $\gamma$. Interestingly, up to the conserved dipolar term $D^a \dot{k}_a$,\footnote{Using the Lie dragging along $k^a|_\gamma$ of $\dot{k}_a \!=\! z \dot{u}_a$ and $D^a = - S^{ab} u_b$ (see Paper I), as well as Killing's equation, the dipolar term $D^a \dot{k}_a$ is easily shown to be a constant of the motion: \beq D^a \ddot{k}_a = D^a (\dot{u}^c \nabla_c k_a) = - (D^a \nabla_a k_c) \dot{u}^c = - \dot{D}_c \dot{k}^c \, . \eeq} the conserved integral \eqref{I} is found to coincide with the Killing energy of the spinning particle, a constant of motion in the dynamics of dipolar particles (see Paper I). Lastly, we can simplify the first term on the right-hand side of \eqref{K_3} by using the equality $k_a|_\scP = z u_a$ and the definition \eqref{defnorms} of the rest mass $m$, and use Leibniz' rule combined with the constraint $D^a u_a = 0$, yielding
\beq\label{K_4}
	I = - mz + \dot{D}^a k_a + \frac{1}{2} S^{ab} \nabla_a k_b \, .
\eeq

Next we turn to the simplification of the formula \eqref{I_1} for $K$. We start by expanding the covariant derivative in the second term by Leibniz' rule. This gives four contributions that we shall consider separately:
\beq\label{I_2}
	K = p^b k^c\delta g_{bc} - (\nabla_d k^a) n_a S^{db} u^{c} \delta g_{bc} - k^a (\nabla_dn_a) S^{db} u^{c} \delta g_{bc} - k^a n_a (\nabla_d\delta g_{bc}) S^{db} u^{c} \, .
\eeq
For the second term, we use Killing's equation $\nabla_{(a} k_{b)} = 0$, as well as $n^a|_\scP = u^a$, so that we can write $(\nabla_d k^a) n_a|_\scP = - z \dot{u}_d$. The third term vanishes, since by Eqs.~\eqref{k=zu} and \eqref{gradient} it is proportional to $u^a \dot{u}_a = 0$. In the
last term, we use $k^a n_a|_\scP = -z$. Renaming some indices and using $k^c|_\scP = z u^c$ in the remaining terms yields
\beq\label{I_3}
	K = (p^a - S^{ab} \dot{u}_b) \delta k_a +  S^{ab} k^c \nabla_a \delta g_{bc} \, ,
\eeq
where we have also used $k^c \delta g_{bc} = \delta k_b$, this last equality coming from Eq.~\eqref{Gauge} with $\xi^a = k^a$. Now let us focus on the first term in the right-hand side of \eqref{I_3}. First we substitute the formula \eqref{p=mu} and simplify the result by using the antisymmetry of $S^{ab}$ and the Leibniz rule, so that
\beq\label{I_4}
	(p^a - S^{ab} \dot{u}_b) \delta k_a = m u^a \delta k_a + \dot{D}^a \delta k_a \, .
\eeq
Second, we use in the right-hand side of \eqref{I_4} the identity $u^a \delta k_a = - 2 \delta z$, which is derived by applying Eqs.~\eqref{Gauge} and \eqref{k=zu} to the equality $u^a k_a = -z$. Substituting all this into \eqref{I_3} gives the following final formula for the integral \eqref{K}:
\beq\label{I_5}
	K = - 2 m \delta z + \dot{D}^a \delta k_a + S^{ab} k^c \nabla_a \delta g_{bc} \, .
\eeq

The formulae \eqref{K_4} and \eqref{I_5} for $I$ and $K$ are consistent with the results established in Sec.~\ref{sec:varid}, for their integral forms \eqref{integrals}. Indeed, the right-hand sides of \eqref{K_4} and \eqref{I_5} are independent of the normal vector $n^a$, and thus of the choice of hypersurface of integration. Moreover, these expressions only involve Lie-dragged quantities: the tensors $k^a,\delta k_a$ and $\delta g_{ab}$ were shown to be Lie-dragged in Sec.~\ref{sec:varid}, while the velocity $u^a$ and the multipoles $(p^a,\dot{D}^a,S^{ab})$ were shown to be Lie-dragged in Paper I, and $z$ and $m$ are constants of motion. Combining all those results with the commutation of the Lie and covariant derivatives, as derived in Paper I, gives, as expected,
\beq
	\dot{I} = z^{-1} \Lik I = 0 \quad \text{and} \quad \dot{K} = z^{-1} \Lik K = 0 \, .
\eeq

\subsection{\texorpdfstring{Linear combination of $I$ and $K$}{}} \label{subsec:combine}

We are finally ready to combine the previous results for $I$ and $K$ in order to compute the quantity $\delta W \equiv -\delta I + K/2$ which appears in the right-hand side of the variational identity \eqref{1st_law}. Combining the variation of Eq.~\eqref{K_4} with \eqref{I_5} readily gives
	\beq
	\delta W = \delta(mz) - \delta(\dot{D}^a k_a) - \frac{1}{2} \delta ( S^a_{\phantom{a}b} \nabla_a k^b ) - m \delta z + \frac{1}{2} \dot{D}^a \delta k_a + \frac{1}{2} S^{ab} k^c \nabla_a \delta g_{bc} \, .
	\eeq
Combining the first and fourth terms yields the monopolar contribution $z \delta m$. By expanding the third term and factorizing by the spin tensor $S^{ab}$ we obtain, after renaming some indices,
	\beq \label{delta Q}
	\delta W = z \delta m - k_a \delta \dot{D}^a - \frac{1}{2} \dot{D}^a \delta k_a - \frac{1}{2} \nabla_a k^b \delta S^a_{\phantom{a}b} - \frac{1}{2} S^{ab} \bigl[ g_{bc} \delta (\nabla_a k^c) - k^c \nabla_a \delta g_{bc} \bigr] \, .
	\eeq
The last step is to show that the last term in the right-hand side vanishes identically. For this we compute the commutator of $\delta$ and $\nabla_a$ applied to both $g_{bc}$ and $k^c$. Then, using \eqref{Gauge} and the condition $\nabla_a g_{bc} = 0$ of metric compatibility, one gets $g_{bc} \delta (\nabla_a k^c) - k^c \nabla_a \delta g_{bc} = k_c \, \delta \Gamma^c_{\phantom{c}ab}$, which is explicitly symmetric. Contracted with the antisymmetric spin $S^{ab}$, this contribution thus vanishes identically. At last, we find
	\beq \label{1st_law_op}
	\delta W = z \delta m - k_a \delta \dot{D}^a - \frac{1}{2} \dot{D}^a \delta k_a - \frac{1}{2} \nabla_a k^b \delta S^a_{\phantom{a}b} \, .
	\eeq

To obtain the first law of binary mechanics in its final form, we note that the result \eqref{1st_law_op} is valid for a single dipolar particle in the binary system. Given the definitions \eqref{integrals} of $I$ and $K$, as well as the linearity of the first law \eqref{1st_law} with respect to the energy-momentum tensor, we have $\delta M - \Omega \, \delta J = \delta W_1 + \delta W_2$, where $\delta W_\ui$ is given by Eq.~\eqref{1st_law_op} and corresponds to the contribution of particle $\ui \in \{ 1,2 \}$ to the binary system. Our final result thus reads
	\beq\label{1st_law_spin}
	\delta M - \Omega \, \delta J = \sum_\ui \Bigl( z_\ui \, \delta m_\ui - \frac{1}{2} \nabla_a k^b \, \delta S_{\ui\,\,b}^a - k_a \, \delta \dot{D}_\ui^a - \frac{1}{2} \dot{D}_\ui^a \, \delta k_a \Bigr) ,
	\eeq
where $z_\ui$, $m_\ui$, $S_{\ui\,\,b}^a$ and $D_\ui^a$ are the redshift, the rest mass, the spin tensor and the mass dipole of the $\ui$-th particle, respectively. Recall that, for each particle, $m_\ui$, $z_\ui = |k|_\ui$ and $\nabla_a k^b$ are all conserved along $\gamma_\ui$, as shown in Paper I.

The variational formula \eqref{1st_law_spin} is one of the most important results of this paper. Let us comment on this particular form of the first law of binary mechanics. First, in the simplest case of a binary system of nonspinning particles, for which $S_\ui^{ab}=0$, which implies $D_\ui^a = 0$, Eq.~\eqref{1st_law_spin} reduces to the standard result already established in Ref.~\cite{Le.al.12}, albeit by following a different route. Second, Eq.~\eqref{1st_law_spin} is \textit{exact} to dipolar order, in the sense that no truncation in the spin tensor has been performed. Third, by imposing the SSC \eqref{SSC}, the last two terms in the right-hand side of \eqref{1st_law_spin} vanish identically, and the first law takes the striking form
\beq\label{1st_law_spin_alt}
	\delta M - \Omega \, \delta J \stackrel{\text{\tiny SSC}}{=} \sum_\ui |k| \, \delta m_\ui - \frac{1}{2} \sum_\ui (\nabla_a k^b) \, \delta S_{\ui\,\,b}^a \, ,
\eeq
where we used the fact that the redshift $z_\ui$ coincides with the norm of the Killing field along the wordline $\gamma_\ui$, and we recall that $m_\ui = - p_\ui^a u^\ui_a$. Equation \eqref{1st_law_spin_alt} naturally suggests that, at higher multipolar order, the right-hand side of the first law may take the form of a multipolar expansion, with multipole index $\ell$, that reads schematically (getting rid of spacetime indices)
\beq\label{1st_law_conj}
	\delta M - \Omega \, \delta J \sim \sum_\ui \sum_{\ell \geqslant 0} \, \underbrace{(\nabla \cdots \nabla k)}_{\ell~\text{derivatives}} \delta Q^{(\ell)}_\ui \, .
\eeq

\section{Spin precession}\label{sec:prec}

In this section, we will focus on a \textit{single} dipolar particle of the binary system and introduce an orthonormal tetrad along its worldline $\gamma$. Combined with the SSC \eqref{SSC}, this will allow us to define an Euclidean 3-vector associated with the covariant spin $S^a$ of the particle. We then discuss the evolution along $\gamma$ of this 3-vector, with respect to a preferred frame that is Lie-dragged along $\gamma$. This will allow us, in the next section, to formulate the first law \eqref{1st_law_spin_alt} in terms of scalar quantities.
\subsection{Orthonormal tetrad} \label{subsec:orthotetrad}
We introduce an orthonormal tetrad $(e_0^a,e_I^a)$, where $e_0^a = u^a$ is taken to coincide with the 4-velocity along $\gamma$, while the uppercase Roman subscript $I \in \{1,2,3\}$ labels the spacelike vectors of the triad $(e_1^a,e_2^a,e_3^a)$. By construction, those four vectors satisfy the orthonormality conditions
	\beq \label{orthotetrad}
	g_{ab} e_I^a e_J^b = \delta_{IJ} \quad \text{and} \quad g_{ab} e_I^a u^b = 0 \, ,
	\eeq
and $u^a u_a = - 1$, where $\delta_{IJ}$ is the Kronecker symbol. We are interested in the evolution of this tetrad along the worldline $\gamma$. It is natural to first expand the vectors $\dot{u}^a$ and $\dot{e}_I^a$ along the tetrad. Using the fact that $\dot{u}^a u_a = 0$ and $ \dot{u}_a e_I^a = - u_a \dot{e}_I^a$, these expansions take the form\footnote{Since the labels $(I,J,K,\dots)$ are but internal Euclidean indices, we may raise and lower them indistinctly.}
	\beq \label{tetradexp}
	\dot{u}^a = a^I e_I^a \quad \text{and} \quad \dot{e}_I^a = a_I u^a + \omega_{IJ} e_J^a \, ,
	\eeq
with the tetrad components $a_I \equiv \dot{u}_a e_I^a = - u_a \dot{e}_I^a$ and $\omega_{IJ} \equiv g_{ab} \dot{e}_I^a e_J^b$. Those are closely related to the so-called Ricci rotation coefficients of the tetrad formalism in general relativity \cite{Wal}. Notice that the orthogonality relations \eqref{orthotetrad} and the metric compatibility $\nabla_c g_{ab} = 0$ imply the antisymmetry of $\omega_{IJ}$:
\beq \label{omegaij}
	\omega_{IJ} = g_{ab} \dot{e}_I^a e_J^b = - \omega_{JI} \, .
\eeq
Consequently, $\omega_{IJ}$ may be viewed as a $3 \times 3$ antisymmetric matrix with 3 degrees of freedom. It is then natural to introduce a dual 3-vector $\bm{\omega} \equiv (\omega^I)$ whose components $\omega^I$ are given by
	\beq \label{omegai}
	\omega^I \equiv -\frac{1}{2} \, \epsilon^{IJK} \omega_{JK} \quad \Longleftrightarrow \quad \omega_{IJ} = - \epsilon_{IJK} \omega^K \, ,
	\eeq
where $\epsilon_{IJK}$ is the totally antisymmetric Levi-Civita symbol, such that $\epsilon_{123} = +1$. Thus far, $\omega_I$ and $a_I$ merely encode the evolution of the tetrad vectors along $\gamma$. As we shall see in the next subsections, for a geometrically motivated class of tetrads, they can be given a fairly simple physical interpretation.
\subsection{Spin precession} \label{subsec:spinprecess}
In Sec.~\ref{subsec:SSC} we showed how the SSC \eqref{SSC} implies the existence of a spacelike spin vector $S^a$, defined in Eq.~\eqref{spin_vector}. Let us now expand this vector over the tetrad. By Eq.~\eqref{ortho_spinvec} we have $S^a u_a = 0$, so the expansion only involves spatial components $S^I$, such that
	\beq \label{expSpinvec}
	S^a = S^I e_I^a \quad \text{with} \quad S_I \equiv S_a e_I^a \, .
	\eeq
An Euclidean spin vector $\mathbf{S}$ can be defined from the three components $S^I$. Those are related to the tetrad components $S_{IJ} \equiv S_{ab} e_I^a e_J^b$ of the spin tensor $S^{ab}$ by an equation analogous to Eq.~\eqref{omegai}, namely
	\beq \label{SiSij}
	S^I = -\frac{1}{2} \, \epsilon^{IJK} S_{JK} \quad \Longleftrightarrow \quad S_{IJ} = - \epsilon_{IJK} S^K\, ,
	\eeq
where we used the definition \eqref{spin_vector} and the formula $\varepsilon_{abcd} u^b e_I^c e_J^d \!=\! \epsilon_{IJK} e^K_a$, which follows from the expansion $\varepsilon^{abcd}|_\gamma = - 4! \, u^{[a} e_1^{b} e_2^{c} e_3^{d]}$ of the volume form on the orthonormal basis $(u^a,e_I^a)$. Equation \eqref{SiSij} allows us to compute the Euclidean norm of the spin vector $\mathbf{S}$, which is found to be conserved, in the sense that [recall Eqs.~\eqref{S} and \eqref{dotS2}]
\beq\label{normspin}
	\delta_{IJ} S^I S^J = \frac{1}{2} S_{ab} S^{ab} = S^2 \, .
\eeq

Next, we look for an equation of evolution along $\gamma$ for the Euclidean spin vector $\mathbf{S}=(S^I)$. To do so, we simply compute the proper time derivative of the equality $S_I = S_a e_I^a$ while using Eqs.~\eqref{fermi-walker} and \eqref{tetradexp}. With the help of Eq.~\eqref{orthotetrad}, the resulting formula can be turned into an evolution equation for $\mathbf{S}$ that reads
	\beq \label{prec_Newt}
	\dot{S}_I = \omega_{IJ} S^J  \quad \Longleftrightarrow \quad \dot{\mathbf{S}} = \bm{\omega} \times \mathbf{S} \, ,
	\eeq
where we used \eqref{omegai} to introduce a cross product. With this Newtonian-looking (but exact) equation of precession for the spin vector $\mathbf{S}$, the vector $\bm{\omega}$ can be interpreted as the precession frequency vector for $\mathbf{S}$. This spin vector precesses in the $(e_I^a)$ frame with an angular frequency $\omega$ given by
\beq \label{normomega}
	\omega^2 \equiv \delta_{IJ} \omega^I \omega^J = \frac{1}{2} \, \omega_{IJ} \omega^{IJ} = \frac{1}{2} \bigl( \dot{e}_I^a \dot{e}^I_a + \dot{u}^a \dot{u}_a \bigr) \, ,
\eeq
where the last two equalities follow from the relations \eqref{omegai} and \eqref{omegaij}, respectively, together with the identity $\delta^{IJ} e_I^a e_J^b = g^{ab} + u^a u^b$. Despite the natural interpretation of \eqref{prec_Newt} as a spin precession equation for the 3-vector $\mathbf{S}$, the precession frequency 3-vector $\bm{\omega}$ depends on the choice of triad [as Eq.~\eqref{normomega} illustrates most clearly], and as such has no invariant meaning.

\subsection{A geometrically-motivated class of tetrads} \label{subsec:liedragtetrad}

In order to give an invariant meaning to the spin precession frequency, thereafter we shall restrict ourselves to the geometrically-motivated class of tetrads $(u^a,e_I^a)$ that are Lie-dragged along the helical Killing field $k^a$, or equivalently along $u^a$. Because we already have $\Lik u^a = 0$ (see Paper I), we additionally require that
\beq\label{Lie_triad}
	\Lik e_I^a = 0 \, .
\eeq
Since $k^a \stackrel{\gamma }{=} z u^a$, the formula \eqref{Lie_triad} implies that $e_I^a$ evolves along $\gamma$ according to $z \dot{e}_I^a = e_I^c \nabla_c k^a$. Using the expansions \eqref{tetradexp} and projecting on the tetrad readily implies
\begin{subequations} \label{omega_et_a}
	\begin{align}
		z \omega_{IJ} &= e_I^a e_J^b \nabla_a k_b \, , \\
		z a_I &= u^a e_I^b \nabla_a k_b \, ,
	\end{align}
\end{subequations}
which shows that $\omega_{IJ}$ is manifestly antisymmetric via Killing's equation. This gives a new interpretation of $\omega_{IJ}$ and $a_I$ which are, up to a factor of $z$, the space-space and space-time components of the Killing 2-form $\nabla_a k_b$ in the Lie-dragged tetrad, respectively.\footnote{The time-time component $u^a u^b \nabla_a k_b$ vanishes identically by virtue of Killing's equation $\nabla_{(a} k_{b)} = 0$.} In other words,
\beq \label{nabkexpand}
	\nabla^a k^b \stackrel{\gamma}{=}  z \omega^{IJ} e_I^a e_J^b + 2 z a^I e_I^{[a} u^{b]}\, .
\eeq
This formula will allow us to compute, in Sec.~\ref{subsec:prec_vort}, the norm of $\nabla_a k_b$ along $\gamma$. Notice also that, since $\Lik \nabla_a k_b = \nabla_a \Lik k_b = 0$ (see Paper I) and $\Lik z = z\dot{z} = 0$, both right-hand sides of Eqs.~\eqref{omega_et_a} are Lie-dragged, such that
	\begin{subequations} \label{zomegadot}
		\begin{align}
			z \dot{\omega}_{IJ} &= \Lik \omega_{IJ} = 0 \, , \\
			z \dot{a}_I &= \Lik a_I = 0 \, .
		\end{align}
	\end{subequations}
Combining this result with the definition \eqref{omegai} gives the important result that the precession frequency vector is constant along the worldline, in this Lie-dragged frame:
	\beq\label{omegadot}
		\dot{\bm{\omega}} = \mathbf{0} \, .
	\eeq
On the other hand, we showed in Paper I that $\Lik S^a = 0$. From the definition \eqref{expSpinvec} and the property \eqref{Lie_triad}, it readily follows that
	\beq \label{dotSi=0}
		z \dot{S}_I = \Lik S_I = 0 \, .
	\eeq
Therefore each component $S^I$ is conserved along the worldline, and so is the vector $\mathbf{S}$. By the equation of spin precession \eqref{prec_Newt}, the spin $\mathbf{S}$ must thus satisfy $\bm{\omega} \times \mathbf{S} = \mathbf{0}$; as a consequence it must be aligned or anti-aligned with the precession frequency vector $\bm{\omega}$. Introducing the notation $\mathbf{n} = (n^I)$ for their common constant direction, such that $\mathbf{n} \cdot \mathbf{n} = 1$, we thus have
\begin{subequations} \label{n}
	\begin{align}
		\bm{\omega} &= \omega \, \mathbf{n} \, , \\
		\mathbf{S} &= S \, \mathbf{n} \, ,
	\end{align}
\end{subequations}
where $\omega$, $S$ and $n_I$ are all constant. We emphasize that although the results \eqref{zomegadot}--\eqref{n} express conservation laws, they have little dynamical contents by themselves, in the sense that they rely crucially on the particular choice of a Lie-dragged triad obeying Eq.~\eqref{Lie_triad}.

Lastly, we mention that taking the unit spin vector $s^a \equiv S^a / S$ as the third leg of a Lie-dragged tetrad $(u^a,e_1^a,e_2^a,s^a)$ allows one to express the formula \eqref{nabkexpand} for the 2-form $\nabla_ak_b$ entirely in terms of the particle's properties. Indeed, combining Eqs.~\eqref{omegai} and \eqref{n} turns Eq.~\eqref{nabkexpand} into
\beq\label{plic}
    \nabla^a k^b \stackrel{\gamma}{=} - 2z \, \bigl( \omega e_1^{[a} e_2^{b]} + u^{[a} \dot{u}^{b]} \bigr) = - 2z \, \bigl( \omega s^{ab} + u^{[a} \dot{u}^{b]} \bigr) \, ,
\eeq
where $s^{ab} \equiv S^{ab} / S$. Indeed, it is easily checked that in such tetrad one has $S^{ab}=S e_1^{[a} e_2^{b]}$. This is a particular case of the general form \eqref{nabkexpand}, in which the only nonvanishing components of the $3 \times 3$ antisymmetric matrix \eqref{omegaij} are $\omega^{12} =  - \omega^{21} = \omega$. Substituting for the formula \eqref{plic} with \eqref{plac} into the equation describing the Lie-dragging along $k^a|_\gamma = zu^a$ of $e_I^a$, which reads $z \dot{e}_I^a = - e^I_b \nabla^a k^b$, yields
\beq\label{edot}
    \dot{e}_1^a + \ui \dot{e}_2^a = \ui \omega (e_1^a + \ui e_2^a)+ a \frac{B_{13}+\ui B_{23}}{1+a\omega} \, u^a \quad \text{and} \quad \dot{e}_3^a = a B_{33} \, u^a \, .
\eeq
Those equations generalize Eqs.~(2.4) of Ref.~\cite{Do.al.15} to nongeodesic motion driven by the spin-coupling to curvature. Comparing with Eqs.~\eqref{tetradexp} yields the values of the tetrad components $a_I$ and $\omega_{IJ}$. Equation \eqref{edot} implies in particular that $\dot{e}_3^a = 0$ if and only if $a B_{33} = 0$, in which case the spin vector $S^a = S e_3^a$ is not only Lie-dragged and Fermi-Walker transported, but also parallel-transported along $\gamma$. This holds for a spinless particle or a test spin ($a=0$) but also if the spacetime has a discrete reflexion symmetry across an equatorial plane ($B_{33} = 0$).

\subsection{A particular Lie-dragged tetrad} \label{subsec:ldt}

\begin{figure}[t!]
    \begin{center}
    	\includegraphics[width=0.4\linewidth]{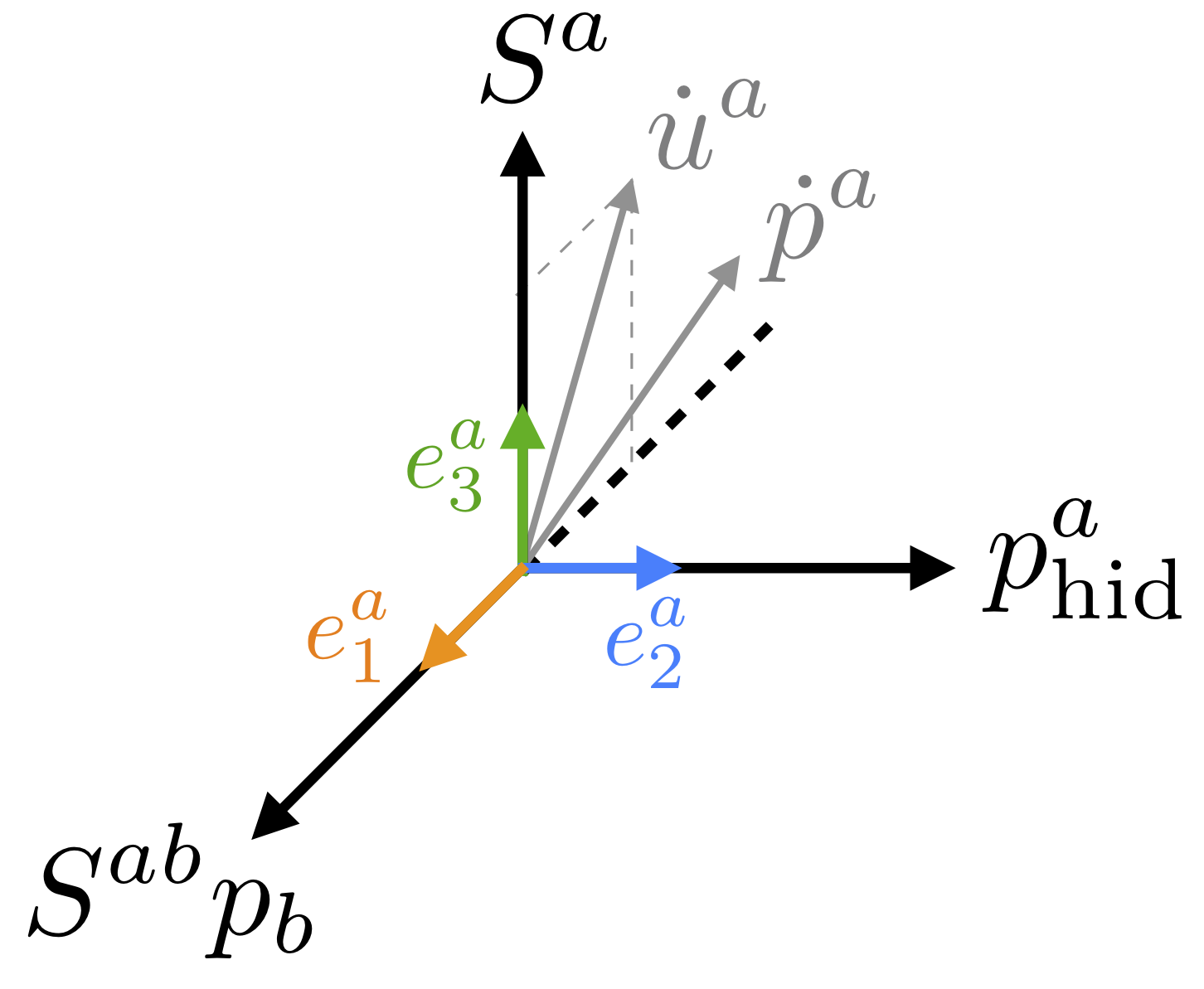}
        \caption{The particular Lie-dragged, orthonormal triad $(e_1^a,e_2^a,e_3^a)$ spanning the local rest frame of the spinning particle, constructed from the particle multipoles and its 4-velocity. The 4-acceleration lies in the $(e_1^a,e_3^a)$ plane, while $\dot{p}^a$ has a non-zero component along the $e_2^a$ direction.}
        \label{fig:tetrad}
    \end{center}
\end{figure}

The legs $e_1^a$ and $e_2^a$ of the Lie-dragged, orthonormal tetrad $(u^a,e_1^a,e_2^a,s^a)$ introduced above were left unspecified. Let us now give an explicit example of an orthonormal tetrad $(u^a,e_I^a)$ in which the legs $e_1^a$ and $e_2^a$ are constructed out of the particle's multipoles $S^{ab},p^a$ and $u^a$, in addition to $e_3^a = s^a$. Consider the following two spacelike vectors:
\beq\label{triad}
    E_1^a \equiv S^{ab}p_b \quad \text{and} \quad E_2^a \equiv S^{ab} \dot{u}_b \, .
\eeq
By construction they are orthogonal to $u^a$ by the SSC \eqref{SSC} and to $e_3^a = s^a$ as a consequence of Eq.~\eqref{spin_vector}. In addition, $E_1^a$ is a linear combination of $\dot{u}^a$ and $S^a$, as shown by Eqs.~\eqref{p=mu_bis} and \eqref{projector}, while $E_2^a$ coincides with the spacelike ``hidden momentum" \eqref{phidden}, as easily seen from Eqs.~\eqref{spin_vector} and \eqref{mumvelo}:
\beq \label{bigEs}
    E_1^a = - S^2 \dot{u}^a + \kappa S^a \quad \text{and} \quad E_2^a = - \varepsilon^a_{\phantom{a}bcd} u^b \dot{u}^c S^d \, .
\eeq
Recall that $\kappa = \tfrac{1}{m} B_{ab} S^a S^b$ vanishes if the metric has a (discrete) reflection symmetry across an equatorial plane, as discussed in Sec.~\ref{subsec:aligned}, in which case $E_1^a = - S^2 \dot{u}^a$. Equation \eqref{bigEs} implies the orthogonality of $E_1^a$ and $E_2^a$. Hence the basis $(u^a,E_1^a,E_2^a,s^a)$ is orthogonal. To obtain an orthonormal tetrad, it suffices to normalize the vectors \eqref{triad}. This is easily done with the help of Eqs.~\eqref{mu-m} and \eqref{p=mu_bis}, which imply
\beq \label{normbigEs}
    |E_1|^2 = S^2 (m^2-\mu^2) \quad \text{and} \quad  |E_2|^2 = m^2 - \mu^2 \, .
\eeq
Defining $e_1^a \equiv E_1^a/|E_1|$ and $e_2^a \equiv E_2^a/|E_2|$, the tetrad $(u^a,e_1^a,e^a_2,s^a)$ is orthonormal. Moreover, having established in Paper I the Lie-dragging along $k^a|_\gamma = zu^a$ of $\dot{u}^a$, $p^a$, $S^a$ and $S^{ab}$, it is clear that this tetrad obeys the property \eqref{Lie_triad}, and is therefore Lie-dragged. This is depicted in Fig.~\ref{fig:tetrad}.

Finally, by inverting the first equation in \eqref{bigEs}, we get the expression of the 4-acceleration within the particular tetrad built from \eqref{triad}, namely
\beq
    S \dot{u}^a = - \sqrt{m^2 - \mu^2} \, e_1^a + \kappa e_3^a \, ,
\eeq
where we used Eq.~\eqref{normbigEs}. Comparing this result to the more general expansion \eqref{plac} allows us to compute some components of $B_{ab}$ with respect to the present tetrad. In particular, we find
\beq
    B_{13} = - \frac{1+a\omega}{a^2} \sqrt{1 - \frac{\mu^2}{m^2}} \quad  \text{and} \quad B_{23} = 0 \, .
\eeq

\subsection{Precession frequency and vorticity}\label{subsec:prec_vort}
In this subsection, we shall express the Euclidean norm $\omega$ of the spin precession frequency vector $\bm{\omega}$ in terms of geometrically-defined quantities related to the helical Killing field \eqref{k}. In particular, the scalar $\omega$ will be shown to be closely related to the vorticity $V^a$ associated to $k^a$. Let us consider the norm $|\nabla k|$ of the 2-form $\nabla_a k_b$ along the worldline $\gamma$. By making use of the orthogonality properties \eqref{orthotetrad}, the formula \eqref{nabkexpand} readily implies
	\beq \label{normnabk}
	|\nabla k|^2 \equiv \frac{1}{2} \nabla^a k^b \nabla_a k_b \stackrel{\gamma}{=} z^2 (\omega^2 - \dot{u}^2) \, ,
	\eeq
where we introduced $\dot{u}^2 \equiv \bm{a} \cdot \bm{a} = \dot{u}^a \dot{u}_a$ by virtue of \eqref{tetradexp}, and we recall that $\omega^2 \equiv \bm{\omega} \cdot \bm{\omega} = \tfrac{1}{2} \omega_{IJ} \omega^{IJ}$. All the scalar fields appearing in this formula are constant along $\gamma$, as was shown in Paper I and Eq.~\eqref{omegadot}. The second term in the right-hand side of Eq.~\eqref{normnabk} can alternatively be written as
    \beq \label{temp}
		z^2 \dot{u}^2 \stackrel{\gamma}{=} \nabla_a |k| \nabla^a |k| \, ,
	\eeq
where we made use of Eq.~(5.6) of Paper I. Combining Eqs.~\eqref{normnabk} and \eqref{temp} yields an exact, coordinate-invariant and frame-independant expression for the (redshifted) norm $z\omega$ of the precession frequency $\bm{\omega}$, namely
\beq\label{covomega}
	z^2 \omega^2 \stackrel{\gamma}{=} \frac{1}{2} \nabla_a k_b \nabla^a k^b + \nabla_a |k| \nabla^a |k| \, .
\eeq
This simple formula generalizes --- for a spinning particle that follows a nongeodesic motion driven by the Mathisson-Papapetrou spin force \eqref{EoM} --- the result of Refs.~\cite{Do.al.14,BiDa3.14,Do.al.15}, which was established in the particular case of a small but massive test spin that follows a geodesic motion in a (properly regularized) helically-symmetric perturbed black hole spacetime.

To linear order in the spin, the motion is not geodesic because $\dot{u}^a = O(S)$, as established in Eq.~\eqref{dotu}. Therefore, the second term in the right-hand side of Eq.~\eqref{covomega} is of quadratic order in the spin, such that
	\beq\label{covomega_lin}
		z \omega \stackrel{\gamma}{=} |\nabla k| + O(S^2) \, .
	\eeq
Comparing this formula with Eq.~(3) of Ref.~\cite{Do.al.14} or Eqs.~(2.11)--(2.12) of Ref.~\cite{BiDa3.14}, we notice an extra factor of the redshift $z$. This can easily be understood because the helical Killing field considered in Ref.~\cite{Do.al.14} is normalized such that $k^a|_\gamma = u^a$ (equivalent to $z = 1$), while the spin precession frequency considered in Ref.~\cite{BiDa3.14} is defined with respect to the coordinate time $t$, and not with respect to the proper time $\tau$ (while $z = \ud \tau / \ud t$ in adapted coordinates).

We now establish a simple relation between the Lorentzian vorticity $V^a$ and the Euclidean spin precession frequency $\bm{\omega}$, through the basis vectors of the Lie-dragged tetrad introduced in Sec.~\ref{subsec:liedragtetrad}. By substititing \eqref{nabkexpand} into the definition \eqref{vorticity} of the vorticity and by using the identity $\varepsilon_{abcd} u^b e_I^c e_J^d = \epsilon_{IJK}e^K_a$ and the definition \eqref{omegai} of the precession 3-vector, we obtain
\beq\label{vorticity_exp}
	V^a = z \omega^I e_I^a \, .
\eeq
Equation \eqref{vorticity_exp} establishes that the components of the vorticity $V^a$ with respect to the Lie-dragged frame $(e_I^a)$ coincide---up to a redshift factor---with the Euclidean components $\omega^I$ of the spin precession frequency vector $\bm{\omega}$. Moreover, by comparing Eq.~\eqref{expSpinvec} and \eqref{vorticity_exp}, we conclude that the Euclidean colinearity \eqref{n} of $\bm{\omega}$ and $\mathbf{S}$ implies the Lorentzian colinearity of $V^a$ and $S^a$, in agreement with the conclusion \eqref{V-S} reached in Sec.~\ref{subsec:helical_binary}. By using the orthonormality of the triad $(e_I^a)$, we find that \eqref{vorticity_exp} implies the following simple relationship between the Lorentzian norm $V$ of the vorticity $V^a$ and the Euclidean norm $\omega$ of the spin precession frequency $\bm{\omega}$:
\beq\label{V=zomega}
    V = z \omega \, .
\eeq
Alternatively, this conclusion could be reached by computing the norm of the vorticity \eqref{vorticity} and comparing it with Eq.~\eqref{covomega}. The conservation of $z$ and $\omega$ along $\gamma$ [recall Eq.~\eqref{omegadot}] is of course compatible with that of $V$, as established in Sec.~\ref{subsec:helical_binary}.

Finally, we come back to the comment we made at the end of Sec.~\ref{subsec:hypersurface}, where we asserted that the spacelike hypersurface $\Sigma$ could not be taken to be globally orthogonal to the helical Killing field $k^a$. Indeed, according to Frobenius' theorem (see, e.g., Ref.~\cite{Wal}), a vector field $\xi^a$ is hypersurface-orthogonal if, and only if, $\xi_{[a} \nabla_b \xi_{c]} = 0$. For the helical Killing vector field \eqref{k}, this is easily shown to imply the vanishing of the twist \eqref{twist}, and consequently of the vorticity \eqref{vorticity}. Because the spin precession vector $\bm{\omega}$ does not vanish, Eqs.~\eqref{vorticity_exp} and \eqref{V=zomega} imply $V^a \neq 0$, and we conclude that $k^a$ cannot be hypersurface-orthogonal.

\section{Hamiltonian first law of mechanics}\label{sec:Hamilton}

In this section, we shall compare our variational formula \eqref{1st_law_spin_alt} to the canonical Hamiltonian first law of mechanics established in Ref.~\cite{Bl.al.13}, for a binary system of spinning particles moving along circular orbits, for spins aligned or anti-aligned with the orbital angular momentum. In Sec.~\ref{subsec:alternative} we first rewrite \eqref{1st_law_spin_alt} in term of the scalars $\omega$ and $S$, to linear order in the spins, by using a Lie-dragged tetrad as introduced in the previous section. Then, in Sec.~\ref{subsec:first_integral} we derive a first integral relation associated with this scalar version of the first law. In Sec.~\ref{subsec:comparison} the scalars $\omega$ and $S$ are related to the Euclidean norms of the canonical spin variable used in Ref.~\cite{Bl.al.13}, and the associated spin precession frequency, which allows us to prove the equivalence of the differential geometric first law \eqref{1st_law_spin_alt} to the Hamiltonian first law of Ref.~\cite{Bl.al.13}. Finally, in Sec.~\ref{subsec:aligned} we prove that the Lie dragging \eqref{Lies} of the spin tensor along the helical Killing field \eqref{k} implies that the canonical spin variable of each spinning particle must be aligned with the orbital angular momentum.

\subsection{Alternative form of the first law}\label{subsec:alternative}

In this first subsection, we write the dipolar contribution $\nabla_a k^b \, \delta S^a_{\phantom{a}b}$ in Eq.~\eqref{1st_law_spin} in terms of the conserved scalars $\omega$ (or $|\nabla k|$) and $\delta S$ that were defined in Eqs.~\eqref{normomega} [or Eq.~\eqref{normnabk}] and \eqref{normspin}, respectively. We will do so at linear order in the spins, since this is all we need in order to compare to the Hamiltonian first law of Ref.~\cite{Bl.al.13}.

By using the formulas \eqref{spin_vector} and \eqref{vort_invert} for $S^a_{\phantom{a}b}$ and $\nabla_a k^b$, as well as the Leibniz rule and the antisymmetry of $\veps^{abcd}$, we readily obtain
\begin{align}\label{DkS1}
    \nabla_ak^b \delta S^a_{\phantom{a}b} &=  \veps^{abcd} \veps_{abef} u^e V^f \delta(u_c S_d) + \veps_{a\phantom{b}ef}^{\phantom{a}b} u^e V^f u_c S_d \, \delta \veps^{a\phantom{b}cd}_{\phantom{a}b} \nonumber \\
    &+ 2\veps^{abcd} k_a \dot{u}_b \, \delta(u_c S_d) + (k_a \dot{u}^b - \dot{u}_a k^b) u_c S_d \, \delta \veps^{a\phantom{b}cd}_{\phantom{a}b} \, .
\end{align}
Let us consider those four terms successively. Using the Leibniz rule and the orthogonality of $u^a$ to both $V^a$ and $S^a$, the first term reduces to $2 V^a \delta S_a - 2 V^c S_c u^a \delta u_a$. With the help of the identity $\delta\veps^{abcd} = - \tfrac{1}{2} \veps^{abcd}g^{ef}\delta g_{ef}$, \vspace{-0.05cm} the second term simplifies to $- V^a S^b \delta g_{ab} + V^c S_c u^a u^b \delta g_{ab}$. Similarly, the third and fourth terms yield $2S^{ab} \dot{k}_a \delta u_b$ and $-S^{ab} \dot{u}_a k^c \delta g_{bc}$, respectively. Using $2 u^a \delta u_a = u^a u^b \delta g_{ab}$, as a consequence of $u^a u_a = - 1$, as well as $\delta k^c = 0$, we then obtain
\beq \label{DkS2}
    \nabla_ak^b \delta S^a_{\phantom{a}b} =  2 V^a \delta S_a - V^a S^b \delta g_{ab} + 2 S^{ab} \dot{k}_a \delta u_b - S^{ab} \dot{u}_a \delta k_b \, .
\eeq

Next, we use the colinearity \eqref{V-S} of $V^a$ and $S^a$, as a consequence of the SSC \eqref{SSC}, so that the first two terms combine to give $2V \delta S$. Using the relation \eqref{V=zomega} we finally obtain the simple result
\beq \label{DkS3}
    \nabla_ak^b \delta S^a_{\phantom{a}b} = 2 z\omega \delta S + \dot{u}_a S^{ab} \delta k_b \, .
\eeq
Therefore, the dipolar contribution in Eq.~\eqref{1st_law_spin_alt} involves a term $z \omega \delta S$ that is linear in spin and a term proportional to $\dot{u}_a S^{ab} \delta k_b$, which is quadratic in spin. Indeed, by virtue of \eqref{dotu} we have
\beq\label{extra_term}
    \dot{u}_a S^{ab} \delta k_b = \frac{1}{m} S^a B_{ab} S^{bc} \delta k_c + O(S^3) \, .
\eeq

We are at last ready to write down the first law of compact binary mechanics in terms of scalar quantities, to \textit{linear} order in the spin amplitudes. To do so, we substitute Eqs.~\eqref{DkS3}--\eqref{extra_term} into \eqref{1st_law_spin_alt} for each particle, and obtain the simple variational formula
\beq\label{1st_law_scalar}
	\delta M - \Omega \, \delta J \stackrel{\text{\tiny SSC}}{=} \sum_\ui z_\ui \left( \delta m_\ui - \omega_\ui \, \delta S_\ui \right) + O(S^2) \, .
\eeq
This is another central result of this paper. Interestingly, for a helically symmetric spacetime that contains one/two black holes, the necessary conditions of vanishing expansion and shear (i.e. Killing horizon) imply that each black hole must be in a state of co-rotation \cite{Fr.al.02,GrLe.13,Bl.al.13,LeGr.18}. By contrast, for a binary system of dipolar particles, the helical isometry merely constrains each spin vector $\mathbf{S}_\ui$ to be aligned with the precession frequency vector $\bm{\omega}_\ui$ [recall Eqs.~\eqref{n}], while the spin amplitude $S_\ui$ of each particle in Eq.~\eqref{1st_law_scalar} is left entirely free.

Now, recalling that $z_\ui = |k|_\ui$ and $z_\ui \omega_\ui = |\nabla k|_\ui + O(S_\ui^2)$, from Paper I and Eq.~\eqref{covomega_lin} above, the variational formula \eqref{1st_law_scalar} looks explicitely like an expansion in powers of (the norms of) the covariant derivatives of the helical Killing vector field $k^a$, namely
\beq\label{1st_law_scalar_bis}
	\delta M - \Omega \, \delta J \stackrel{\text{\tiny SSC}}{=} \sum_\ui \left( |k|_\ui \, \delta m_\ui - |\nabla k|_\ui \, \delta S_\ui \right) + O(S^2) \, .
\eeq
This naturally suggests that, at the next quadrupolar order, one might obtain an additional contribution of the form $\sum_\ui |\nabla\nabla k|_\ui \, \delta Q_\ui$, where the double covariant derivative of the helical Killing field can be related to the curvature tensor through the Kostant formula (Paper I), and $Q_\ui$ would be the spacetime norm of the quadrupole moment tensor of each particle.

\subsection{First integral relationship}\label{subsec:first_integral}

As shown e.g. in Refs.~\cite{Le.al.12,Bl.al.13,GrLe.13,Le.14,Le.15,BlLe.17}, each variational first law of binary mechanics implies the existence of an associated first integral relationship. By applying Euler's theorem to the homogeneous function of degree one $M(\sqrt{J},\sqrt{S_1},\sqrt{S_2},m_1,m_2)$, the so-called first integral associated with the variational formula \eqref{1st_law_scalar} simply reads
\beq\label{1st_int_spin}
	M - 2\Omega J \stackrel{\text{\tiny SSC}}{=} \sum_\ui z_\ui \left( m_\ui - 2 \omega_\ui S_\ui \right) + O(S^2) \, .
\eeq
A closely related algebraic formula can be derived from the Komar-type notions of mass and angular momentum. Indeed, as established in App.~\ref{app:Komar} the relevant linear combination of the Komar mass $M_\text{K}$ and angular momentum $J_\text{K}$ is precisely given, at dipolar order, by
\beq\label{1st_int_Komar}
	M_\text{K} - 2 \Omega J_\text{K} = - \sum_\ui \bigl( p^a_\ui k_a + 2 S_\ui^a V_a + D^a_\ui \dot{k}_a \bigr) \, .
\eeq
Notice the couplings of the particle's multipoles to the Killing field and its derivatives in the right-hand side. After imposing the SSC \eqref{SSC} and using $p_\ui^a k_a = - m_\ui z_\ui$ and $S_\ui^a V_a = S_\ui V = z_\ui \omega_\ui S_\ui$, as a consequence of Eqs.~\eqref{k=zu}, \eqref{m}, \eqref{V-S} and \eqref{vorticity_exp}, we readily obtain the simple algebraic formula
\beq\label{1st_int_Komar_bis}
	M_\text{K} - 2 \Omega J_\text{K} \stackrel{\text{\tiny SSC}}{=} \sum_\ui z_\ui \left( m_\ui - 2 \omega_\ui S_\ui \right) .
\eeq
Assuming that our helically symmetric spacetimes would obey appropriate falloff conditions \cite{Sh.al.04,Go.07}, it can be shown that $M_\text{K} = M$ and $J_\text{K} = J$, so that the Komar-type derivation of the first integral relation is consistent with the formula \eqref{1st_int_spin}. In fact, the algebraic formula \eqref{1st_int_Komar_bis} suggests that the first integral \eqref{1st_int_spin} is \textit{exact} at dipolar order, and not merely valid to linear order in the spin amplitudes $S_\ui$.

Moreover, as shown in App.~\ref{app:Komar} the formula \eqref{1st_int_Komar} can alternatively be written in the form
\beq\label{1st_int_Komar_ter}
	M_\text{K} - 2 \Omega J_\text{K} \stackrel{\text{\tiny SSC}}{=} - \sum_\ui \bigl( p^a_\ui k_a + S_\ui^{ab} \nabla_a k_b \bigr) = - \sum_\ui \bigl[ E^{(p)}_\ui + 2 E^{(S)}_\ui \bigr] \, ,
\eeq
where we imposed once again the SSC \eqref{SSC} and defined $E^{(p)}_\ui \equiv p_\ui^a k_a$ and $E^{(S)}_\ui \equiv S_\ui^{ab} \nabla_a k_b / 2$. The right-hand side of Eq.~\eqref{1st_int_Komar_ter} is closely related to the sum of the Killing energies $E_\ui = E^{(p)}_\ui + E^{(S)}_\ui$ of the spinning particles. We established in Paper I that, for each particle, the monopolar and dipolar contributions $E^{(p)}_\ui$ and $E^{(S)}_\ui$ to $E_\ui$ are separately conserved. This is consistent with the fact that $M_\text{K}$, $J_\text{K}$ and $\Omega$ are constants.

\subsection{Comparison to the Hamiltonian first law}\label{subsec:comparison}

By using the canonical Arnowitt-Deser-Misner (ADM) Hamiltonian framework of general relativity applied to spinning point particles, the authors of Ref.~\cite{Bl.al.13} derived a first law of mechanics for binary systems of compacts objects with spins aligned (or anti-aligned) with the orbital angular momentum, to \textit{linear} order in the spins. Our goal is to relate this earlier result to the scalar version \eqref{1st_law_scalar} of the first law, which also holds to linear order in spins.

In Sec.~\ref{sec:prec} we have established the geometrical precession, with respect to an orthonormal frame $(e_I^a)$ orthogonal to $\gamma$ and Lie-dragged along it, of a Euclidean spin vector $\mathbf{S}$ orthogonal to $\gamma$. Using $z = \ud \tau / \ud t$, which holds in any coordinate system adapted to the helical isometry (see Paper I), Eqs.~\eqref{prec_Newt} and \eqref{covomega_lin} can be rewritten, for each particle, in the equivalent form
\beq\label{prec_S}
	\frac{\ud \mathbf{S}_\ui}{\ud t} = z_\ui \bm{\omega}_\ui \times \mathbf{S}_\ui \, , \quad \text{where} \quad z_\ui \omega_\ui \stackrel{\gamma_\ui}{=} |\nabla k|_\ui + O(S_\ui^2) \, .
\eeq
As shown in Refs.~\cite{Da.al.08,BiDa3.14}, one can relate this well established kinematical result to \textit{dynamical} properties of spin-orbit coupling in a binary system of spinning compact objects.

This can be done in the context of the canonical ADM Hamiltonian framework of general relativity, applied to a binary system of spinning point masses with canonical positions $\mathbf{r}_\ui(t)$, momenta $\mathbf{p}_\ui(t)$ and spins $\bar{\mathbf{S}}_\ui(t)$. In the center-of-mass frame, the dynamics depends on the relative position $\mathbf{r} \equiv \mathbf{r}_1 - \mathbf{r}_2$, the relative momentum $\mathbf{p} \equiv \mathbf{p}_1 = - \mathbf{p}_2$ and the individual spins. The evolution of the canonical variables is then governed, to linear order in the spins, by a canonical Hamiltonian
\beq\label{H}
	H(\mathbf{r},\mathbf{p},\bar{\mathbf{S}}_\ui) = H_\text{orb}(\mathbf{r},\mathbf{p}) + \sum_\ui \mathbf{\Omega}_\ui(\mathbf{r},\mathbf{p}) \cdot \bar{\mathbf{S}}_\ui \, ,
\eeq
where the pseudo-vectors $\mathbf{\Omega}_\ui = \sigma_\ui \, \mathbf{L}$ are both proportional to the orbital angular momentum $\mathbf{L} = \mathbf{r} \times \mathbf{p}$, while $\sigma_\ui$ is closely related to the gyro-gravitomagnetic ratio of particle \# $\ui$ \cite{Da.al2.08}. The usual Poisson bracket structure of the Cartesian components $\bar{S}_\ui^j$ of the canonical spin $\bar{\mathbf{S}}_\ui$ ensures that the spin-orbit (linear-in-spin) part of the canonical Hamiltonian \eqref{H} implies Newtonian-looking, but exact precession equations of the form \cite{Da.al.08}
\beq\label{prec_Scan}
	\frac{\ud \bar{\mathbf{S}}_\ui}{\ud t} = \mathbf{\Omega}_\ui \times \bar{\mathbf{S}}_\ui \, .
\eeq
Consequently, the Euclidean norm $\Vert \bar{\mathbf{S}}_\ui \Vert$ of each canonical spin variable is conserved. For that reason, the canonical variables $\bar{\mathbf{S}}_\ui$ are known as the ``constant-in-magnitude'' spins. Those variables are, however, by no means unique. In particular, it can be shown that the gauge freedom (local rotation group) to perform an infinitesimal rotation of $\bar{\mathbf{S}}_\ui$ can be seen as being induced by an infinitesimal canonical transformation in the full phase space \cite{Da.al.08}.

Despite the stricking similarity between Eqs.~\eqref{prec_S} and \eqref{prec_Scan}, the canonical spin variable $\bar{\mathbf{S}}_\ui$ needs not coincide with the Euclidean spin vector $\mathbf{S}_\ui$ constructed from the components of the 4-vector $S_\ui^a$ along an orthonormal triad $(e^a_I)$, and the precession frequency $z_\ui \bm{\omega}_\ui$ needs not coincide with the coefficient $\mathbf{\Omega}_\ui$ appearing in the spin-orbit piece of the canonical Hamiltonian \eqref{H}. In fact, several definitions of a globally $\text{SO}(3)$-compatible, canonical spin variable $\bar{\mathbf{S}}$ constructed from a spacelike, covariant 4-vector $S^a$ are possible \cite{Da.al.08,BiDa3.14}. For instance, given the spatial components $S_i$ of the covector $S_a = g_{ab} S^b$ with respect to an ADM coordinate system compatible with the Hamiltonian formulation, one may define a particular canonical spin variable according to
\beq\label{Scan}
	\bar{S}^i \equiv H^{ij} S_j \, ,
\eeq
where $H^{ij}$ is the \textit{unique}, symmetric and positive definite square-root of the effective metric $G^{ij} \equiv g^{ij} - 2 g^{0(i} v^{j)} + g^{00} v^i v^j$, which is constructed from the components $g^{\alpha\beta}$ of the inverse metric and the components $v^\alpha = u^\alpha / u^0$ of the coordinate 3-velocity of the particle \cite{Da.al.08,Bo.al.13}. The key point is that the Euclidean norm $\bar{S} \equiv \Vert \bar{\mathbf{S}} \Vert$ of the canonical spin variable \eqref{Scan} is numerically equal to the norm \eqref{normspin} of the spin vector $\mathbf{S}$ (and is thus conserved):
\beq\label{spin-norms}
	\bar{S}^2 \equiv \delta_{ij} \bar{S}^i \bar{S}^j = S^2 \, .
\eeq

Moreover, the global $\text{SO}(3)$ symmetry of the Hamiltonian dynamics generated by \eqref{H} implies, in particular, that the conserved total angular momentum vector $\mathbf{J}$ has the simple, additive form $\mathbf{J} = \mathbf{L} + \sum_\ui \bar{\mathbf{S}}_\ui$ \cite{Da.al.08}. For spins aligned or anti-aligned with the orbital angular momentum $\mathbf{L}$, this reduces to the algebraic equality
\beq\label{J-L-S}
	J = L + \sum_\ui \bar{S}_\ui \, .
\eeq
Finally, in the particular case of circular orbits of angular velocity $\Omega$, the analysis of Ref.~\cite{BiDa3.14} (see also \cite{Do.al.14}) shows that, for each particle, the Euclidean norm $\Omega_\ui \equiv \Vert \mathbf{\Omega}_\ui \Vert$ of the precession frequency vector of the canonical spin variable $\bar{\mathbf{S}}_\ui$ is related to the Lorentzian norm \eqref{normnabk} of the helical Killing 2-form $\nabla_a k_b$ by the simple numerical link
\beq\label{precession-freq}
	\Omega_\ui = \Omega - \vert \nabla k \vert_\ui \, .
\eeq
The occurrence of the circular-orbit angular velocity $\Omega$ can be understood from the fact that the precession frequency $\Omega_\ui$ is defined with respect to the coordinate time of the global chart associated with the 3+1 split required to construct the canonical Hamiltonian $\eqref{H}$, while the (redshifted) precession frequency $z_\ui \omega_\ui = |\nabla k|_\ui$ is defined with respect to a local spatial frame that is Lie-dragged along the spinning particle's worldline. Importantly, the relation \eqref{precession-freq} was established to linear order in the spin amplitudes $S_\ui$.

Substituting the relationships \eqref{spin-norms}--\eqref{precession-freq} into the geometric, scalar first law \eqref{1st_law_scalar}, we readily find
\beq\label{first_law_Hamil}
	\delta M - \Omega \, \delta L = \sum_\ui (z_\ui \, \delta m_\ui + \Omega_\ui \, \delta \bar{S}_\ui) + O(\bar{S}^2) \, .
\eeq
Provided that the total mass $M$ coincides with the on-shell value of the Hamiltonian \eqref{H}, the variational formula \eqref{first_law_Hamil} precisely agrees with the Hamiltonian first law of mechanics derived in Ref.~\cite{Bl.al.13} [see Eq~(4.6) there], for binary systems of spinning compact objects moving along circular orbits, with spins aligned or anti-aligned with the orbital angular momentum, to linear order in the spins. The first integral associated with the variational formula \eqref{first_law_Hamil} reads \cite{Bl.al.13}
\beq\label{first_integral_Hamil}
	M - 2 \Omega L = \sum_\ui (z_\ui m_\ui + 2\Omega_\ui \bar{S}_\ui) + O(\bar{S}^2) \, .
\eeq
Combined with Eqs.~\eqref{J-L-S} and \eqref{precession-freq}, together with the formula \eqref{covomega_lin}, this agrees with the ``geometrical'' first integral relation \eqref{1st_int_spin}.

\subsection{Discussion on spin precession}\label{subsec:aligned}

For a binary system of spinning particles on a circular orbit, one expects that the canonical spins $\bar{\mathbf{S}}_\ui$ be aligned or anti-aligned with the orbital angular momentum $\mathbf{L} = \mathbf{r} \times \mathbf{p}$, otherwise the canonical spins $\bar{\mathbf{S}}_\ui$ and the instantaneous orbital plane (orthogonal to $\mathbf{L}$) would precess, arguably breaking the helical isometry. In our non-Hamiltonian, isometric context, a natural question is whether the Lie-dragging of the spin covector $S_a$ of each spinning particle, as established in Paper I, implies that the associated canonical spin variable $\bar{\mathbf{S}}$ must be aligned or anti-aligned with the orbital angular momentum $\mathbf{L}$. For each particle, the spin precession equation \eqref{prec_Scan} for the canonical spin variable $\bar{\mathbf{S}}$ can be rewritten as
\beq\label{plout}
    (\mathbf{\Omega} \times \bar{\mathbf{S}})_i = \frac{\ud \bar{S}_i}{\ud t} = N \frac{\ud \bar{S}_i}{\ud \tau} = N z^{-1} \Lik \bar{S}_i \, ,
\eeq
where we used the 3+1 formula $\ud\tau = N \ud t$ in the second equality, with $N$ the lapse function, and the colinearity condition \eqref{k=zu} in the last equality. The construction of a canonical spin variable $\bar{\mathbf{S}} = (\bar{S}_i)$ presented in \cite{Da.al.08} [recall Eq.~\eqref{Scan}] was later reformulated in \cite{Bo.al.13} in terms of a specific orthonormal tetrad $(u^a,\epsilon_I^a)$, such that $\bar{S}_I = \epsilon_I^a S_a$ coincides numerically with $\bar{S}_i$. The relation \eqref{plout} can then be reformulated as
\beq
    (\mathbf{\Omega} \times \bar{\mathbf{S}})_i = N z^{-1} (\Lik \epsilon_I^a) S_a \, ,
\eeq
where we used $\Lik S_a \!=\! 0$, as shown in Paper I. Therefore, the canonical spin variable $\bar{\mathbf{S}} = (\bar{S}_i)$ is aligned (or anti-aligned) with the spin precession frequency $\mathbf{\Omega}$, and consequently with the orbital angular momentum $\mathbf{L} \propto \mathbf{\Omega}$ if, and only if, the triad $(\epsilon^a_I)$ used to define $\bar{\mathbf{S}}$ is Lie-dragged along $\gamma$. In other words, a necessary and sufficient condition for no spin precession nor orbital plane precession is the requirement that the tetrad $(u^a,\epsilon_I^a)$ of Ref.~\cite{Bo.al.13} belongs to the class of geometrically-motivated tetrads obeying the evolution law \eqref{Lie_triad} along $\gamma$. Under such a condition, the orbital plane coincides with an equatorial plane, with respect to which the metric has a reflection symmetry; that is, there is a discrete symmetry under a coordinate transformation of the form $\theta \to \pi - \theta$, for some polar angle $\theta$. One may then borrow the analysis of Ref.~\cite{Do.al.15}, to show that
\beq\label{BSS_bis}
    B_{ab} S^a S^b = 0 \, .
\eeq
Assuming Eq.~\eqref{BSS_bis}, we showed in Sec.~\ref{subsec:SSC} that the spin vector $S^a$ is parallel transported along $\gamma$ [recall Eq.~\eqref{FWv2}], and argued in Sec.~\ref{subsubsec:reflexion} that the 4-acceleration is orthogonal to both the 4-velocity and the spin vector, as expected in the case of circular motion.

Now we come back to the question of whether the ADM-square-root-triad is Lie-dragged along the helical Killing vector $k^a$. First, let us recall the construction of this triad, as presented in \cite{Bo.al.13}. Given a coordinate system $(x^\alpha)$, compute the spatial (covariant) components $\gamma_{ij}$ of the projection tensor $\gamma_{ab}=g_{ab} + u_a u_b$  with respect to those coordinates, and define the spatial components $(\epsilon_I)_i$ of the covector $(\epsilon_I)_a=g_{ab}\epsilon_I^b$ by requiring that\footnote{The timelike component $(\epsilon_I)_0$ is determined by the orthogonality condition $u^a (\epsilon_I)_a = 0$, which implies $(\epsilon_I)_0 = - v^i (\epsilon_I)_i$, with $v^i$ the coordinate 3-velocity of the particle.}
\beq
    \delta^{IJ} (\epsilon_I)_i (\epsilon_J)_j = \gamma_{ij} \qquad \text{and} \qquad (\epsilon_I)_i = (\epsilon_i)_I \, .
\eeq
While the first condition is satisfied by any orthonormal tetrad $(u^a,\epsilon_I^a)$, the second condition (symmetry under the exchange $i\leftrightarrow I$) defines a \textit{unique} triad $(\epsilon_I^a)$ by requiring that the $3 \times 3$ matrix $M$ with components $m_{ij}=(\epsilon_i)_j$ is the unique, positive-definite square-root of the $3 \times 3$ matrix $\Gamma$ of components $\gamma_{ij}$, i.e., such that $M^2=\Gamma$.

It is important to note that this ``square-root triad'' is not defined covariantly, but rather by specifying its coordinate components with respect to an \textit{a priori} given coordinate system. For example, the authors of \cite{Bo.al.13} construct such a ``square-root triad" using harmonic coordinates. Using different coordinates for this construction---for instance ADM coordinates---will generally define a different triad. Yet, regardless of the coordinate choice, from a geometrical point of view, such a triad is defined from the metric and particle 4-velocity, allowing us to expect that it would be Lie-dragged in presence of a helical Killing vector $k^a$.

While we were not able to show that the ADM-coordinate square-root triad of Ref.~\cite{Da.al.08,Bo.al.13} is Lie-dragged along $k^a|_\gamma$, we can construct a square-root triad that is Lie-dragged, by using coordinates adapted to the helical symmetry. This involves a Fermi normal coordinate (FNC) system $(t,x^i)$ erected along the worldline $\gamma$ of either of the two particles; see e.g. \cite{Po.al.11} for details. With respect to those coordinates, we have $g_{ij}=\delta_{ij}$ and $u_i=0$ along $\gamma$, implying $\gamma_{ij}=\delta_{ij}$ along $\gamma$. The square-root triad is thus made of the components of a matrix whose square is the identity matrix $I_3$, namely  $I_3$ itself. In other words, the components of the square-root triad in FNC are simply $(\epsilon_I)_i = \delta_{Ii}$ and $(\epsilon_I)_0 = -v^I = 0$ given $u^i \!=\! 0$. By construction, in FNC the only coordinate that varies along $\gamma$ is $t$. But $\gamma$ is also an integral curve of $k^a$. Therefore, the FNC system $(t,x^i)$ is ``adapted'' to the helical symmetry, in the sense of Ref.~\cite{Wal} (see Sec.~C.2 there). Consequently, the FNC components of $\Lik (\epsilon_I)_a$ are given by $\Lik (\epsilon_I)_i = \partial_t \delta_{Ii} = 0$ and $\Lik (\epsilon_I)_0 = 0$. We conclude that this specific triad is Lie-dragged along $\gamma$.

\acknowledgments

The authors acknowledge the financial support of the Action F\'ed\'eratrice PhyFOG and of the Scientific Council of the Paris Observatory. ALT acknowledges the hospitality of the Brazilian Center for Research in Physics (CBPF), where part of this work was carried out, and is most grateful to Luc Blanchet for allowing the authors to adapt in App.\ref{app:fluid} the contents of common research notes that date back to 2012. PR wishes to thank Lorenzo Rossi for helpful comments on the computations involved in Sec.~\ref{sec:varid}.

\appendix

\section{Summary of conventions and notations}\label{app:conventions}

Our sign conventions are those of \cite{Wal}. In particular, the metric signature is $(-,+,+,+)$, the Riemann tensor satisfies $2\nabla_{[a} \nabla_{b]} \omega_c = R_{abc}^{\phantom{abc}d} \omega_d$ for any 1-form $\omega_a$, and the Ricci tensor is defined by $R_{ab} = R_{acb}^{\phantom{acb}c}$. Abstract indices are denoted using letters $(a,b,c,\dots)$ from the beginning of the Latin alphabet, tensor components with respect to a given basis are denoted by Greek letters $(\alpha,\beta,\gamma,\dots)$, and indices $(i,j,k,\dots)$ from the middle of the Latin alphabet refer to purely spatial components. Uppercase letters $(I,J,K,\dots)$ denote internal Lorentz indices associated with an orthonormal triad $(e^a_I)$, with $I \in \{1,2,3\}$, and the Roman font subscript $\ui \in \{1,2\}$ is used to denote the two particles of the binary. We use interchangeably the notations $|_\gamma$ and $\stackrel{\gamma}{=}$ for an equality that holds only along the worldline $\gamma$ of a particle. Boldface symbols denote differential forms in Sec.~\ref{sec:varid} only, and Euclidean 3-vectors in Secs.~\ref{sec:prec} and \ref{sec:Hamilton}. Throughout
this paper, we use geometrized units such that $G = c = 1$. For the convenience of the reader, a list of the symbols used most frequently is given in Tab.~\ref{Table}.

\begin{table}[!ht]
    \caption{List of frequently used symbols.}
    \vspace{0.2cm}
	\begin{tabular}{cc}
		\toprule
		\textbf{Symbol}         & \textbf{Description} \\
		\midrule
		\textbf{Sets}           &                                \\
		$\mathcal{M}$           & spacetime manifold             \\
        $\Sigma$                & spacelike hypersurface         \\
		$\gamma$                & particle worldline             \\
		$\scP$                  & point $\Sigma \cap \gamma$     \\
		\midrule
		\textbf{Geometry}       &                                \\
		$g_{ab}$                & metric on $\mathcal{M}$        \\
        $\nabla_a$              & covariant derivative           \\
		$\varepsilon_{abcd}$    & canonical volume form          \\
		$R_{abcd}$              & Riemann curvature tensor       \\
        $Q_{ab}$                & Noether charge 2-form          \\
		$\xi^a$                 & generic Killing vector         \\
		$k^a$                   & helical Killing vector         \\
		$\varpi^a$              & twist associated with $k^a$    \\
        $\Lik$                  & Lie derivative along $k^a$     \\
        $n^a$                   & unit normal to $\Sigma$        \\
        $e_I^a$                 & generic triad                  \\
        $\epsilon_I^a$          & ``square-root'' triad          \\
		\midrule
		\textbf{Particles}      &                                \\
		$\tau$                  & proper time                    \\
		$m$                     & rest mass                      \\
		$\mu$                   & dynamical mass                 \\
		$z$                     & redshift parameter             \\
		$\omega$                & precession frequency           \\
        $u^a$                   & 4-velocity                     \\
        $\dot{u}^a$             & 4-acceleration                 \\
		$p^a$                   & 4-momentum                     \\
		$S^a$                   & spin vector                    \\
		$D^a$                   & mass dipole                    \\
        $V^a$                   & vorticity                      \\
        $S^{ab}$                & spin tensor            	     \\
		$T^{ab}$                & energy-momentum tensor         \\
		$\delta_4$              & invariant Dirac distribution   \\
        \bottomrule
	\end{tabular}
    \label{Table}
\end{table}

\section{Generalized first law of Friedman, Ury{\=u} and Shibata}\label{app:fluid}

Equation \eqref{bim!} relates the \textit{Eulerian} variations $\delta \mathcal{Q}$ of various quantities $\mathcal{Q}$, meaning their changes at a fixed point (same coordinate values in both spacetimes). However, small changes in the matter fields are more naturally described in terms of the \textit{Lagrangian} variations $\Delta \mathcal{Q}$, corresponding to the variation of $\mathcal{Q}$ while following the fluid element as it is displaced. The link between the Eulerian and Lagrangian variations can formally be written as
\beq\label{Euler-Lagrange}
	\Delta = \delta + \Lila \, ,
\eeq
where $\Lila$ is the Lie derivative along the Lagrangian displacement $\lambda^a$ connecting a fluid element in the unperturbed spacetime $g_{ab}$ to the corresponding fluid element in the perturbed spacetime $g_{ab} + \delta g_{ab}$. An important property is the existence of a class of trivial displacements leaving the Lagrangian perturbations $\Delta \mathcal{Q}$ unchanged.\footnote{These include Lagrangian displacements of the form $\lambda^a = f u^a$, where $f$ is an arbitrary scalar field and $u^a$ is the 4-velocity field of the fluid. Such displacements merely ``push'' the fluid along its own flow.} These can be used to freely specify the timelike part of $\lambda^a$. Hereafter we shall consider only purely spatial Lagrangian displacements, in the sense that $\lambda^a u_a = 0$, with $u^a$ the 4-velocity of the fluid. See Refs.~\cite{Ta.69,Ca.73,FrSc.75,ScSo.77,Fr.78} for more details on Eulerian and Lagrangian variations in relativistic fluids.

Replacing the Eulerian variations $\delta$ by Lagrangian variations $\Delta$ in the RHS of Eq.~\eqref{bim!}, and remembering the gauge choice $\delta \xi^a = 0$ in \eqref{Gauge}, the variational formula can be recast in the form
\beq\label{1st_law_Delta}
	\delta H_\xi = - \int_{\Sigma} \Delta \bigl( \ud \Sigma_a T^a_{\phantom{a}b} \bigr) \xi^b + \frac{1}{2} \int_\Sigma \ud \Sigma_a \xi^a \, T^{bc} \Delta g_{bc} \, .
\eeq
Indeed, the difference between the right-hand sides of \eqref{bim!} and \eqref{1st_law_Delta} can easily be shown to vanish. We have proven in Paper I that the matter source must satisfy $\Lixi T^{ab} = 0$. This constraint, together with $\Lila g_{ab} = 2 \nabla_{(a} \lambda_{b)}$ and $\Delta \xi^a = \Lila \xi^a = - \Lixi \lambda^a$, as well as Stokes' theorem, can be used to write the difference between Eqs.~\eqref{bim!} and \eqref{1st_law_Delta} as a surface integral, namely
\beq
    \int_\Sigma \Lila \bigl( \ud \Sigma_a T^a_{\phantom{a}b} \bigr) \xi^b - \frac{1}{2} \int_\Sigma \ud \Sigma_a \xi^a \, T^{bc} \Lila g_{bc} = \oint_S \ud S_\alpha \, B^\alpha \, ,
\eeq
where $B^\alpha = \lambda^\alpha \, T^{0\beta} \xi_\beta + \xi^\alpha \, T^{0\beta} \lambda_\beta - T^{\alpha \beta} \lambda_\beta$. This surface integral over the topological 2-sphere $S$ vanishes because the energy-momentum tensor appearing in $B^\alpha$ has a compact support.

The generalized first law \eqref{1st_law} is valid for an arbitrary combination of ``matter'' sources described by the energy-momentum tensor $T^{ab}$. We shall now apply it to the case of a perfect fluid matter source. Consider a distribution of matter made of, say, two well-separated balls of perfect fluid modelling two orbiting bodies, and described by the energy-momentum tensor
\beq\label{Tab}
    T_{ab} = \left(\epsilon + p\right) u_a u_b + p\,g_{ab} \, .
\eeq
The 4-velocity $u^a$ of the fluid is normalized according to $g_{ab} u^a u^b = -1$. Its equation of state is given in the form of the proper energy density $\epsilon(\rho,s)$ and pressure $p(\rho,s)$ as functions of the entropy per unit baryonic mass $s$ and proper mass density $\rho$, such that the mass current $j^a = \rho u^a$ is conserved, i.e., $\nabla_a j^a = 0$. It also proves convenient to decompose the redshifted 4-velocity as \cite{Fr.al.02,Ur.al.10}
\beq\label{zu}
	z u^a = \xi^a + w^a \, , \quad \text{with} \quad w^a \xi_a = 0 \, ,
\eeq
where $z \equiv - \xi^a u_a$ is the redshift factor. Indeed, the integral curves of the Killing vector field $\xi^a$ define some preferred worldlines that can be seen as the main stream velocity of the fluid, while the spacelike velocity field $w^a$ measures the ``internal'' velocity of the fluid elements with respect to that flow, i.e., with respect to a frame comoving with $\xi^a$.

While comparing two nearby fluid configurations, for each fluid particle the Lagrangian variations of the proper energy density $\epsilon$, proper mass density $\rho$, and specific entropy $s$ must be related through the usual first law of thermodynamics,
\beq\label{1st_law_thermo}
	\Delta \epsilon = H \, \Delta \rho + T \rho \, \Delta s \, ,
\eeq
where $H \equiv (\epsilon+p)/\rho$ is the specific enthalpy and $T$ the temperature of the fluid. Notice that Eq.~\eqref{1st_law_thermo} presupposes a ``physical process'' \textit{interpretation} of the generalized first law \eqref{1st_law}--- rather than a comparison between two distinct spacetimes admitting an isometry---whereby the fluid is assumed to evolve from one quasi-equilibrium configuration to another under the effect of gravitational radiation reaction.\footnote{This is analogous to the physical process version \cite{Wal2} of the first law of black hole thermodynamics \cite{Ba.al.73}, relating the variations in mass, angular momentum, and surface area between two nearby stationary and axisymmetric solutions, as a result of a physical perturbation (e.g. a small mass plunging into the hole).}

The fact that $\Delta$ maps fluid trajectories to fluid trajectories, together with the normalization $u^a u_a = -1$ of the 4-velocity, implies $\Delta u^a = \frac{1}{2} \, u^a u^b u^c \Delta g_{bc}$ \cite{Ca.73,FrSc.75,Fr.78}, from which we easily deduce $\Delta(z u^a) = 0$, as well as $\Delta z / z = u_a \Delta u^a$. We also have $g^{ab} \Delta g_{ab} = 2 \Delta\ln{\sqrt{-g}}$. The previous formulas, together with the decomposition \eqref{zu}, can be used to express the two integrants in the right-hand side of Eq.~\eqref{1st_law_Delta} as
\begin{subequations}
	\begin{align}
		\Delta(- \ud \Sigma_a T^a_{\phantom{a}b}) \xi^b &= \Delta(\ud^3 x \, \sqrt{-g} \, \epsilon) + w^a \Delta(H u_a \ud M_\text{bar}) \, , \\
		\frac{1}{2} \, \ud \Sigma_a \xi^a \, T^{bc} \Delta g_{bc} &= \ud^3 x \, p \, \Delta \sqrt{-g} - \ud M_\text{bar} H \, \Delta z \, ,
	\end{align}
\end{subequations}
where $\ud M_\text{bar} \equiv j^a \ud \Sigma_a$ is the baryonic mass element, also given by $\ud M_\text{bar} = \rho_* \, \ud^3 x$, where $\rho_* \equiv \sqrt{-g} \, u^0 \rho$ is the so-called ``coordinate'' mass density.\footnote{The covariant conservation law $\nabla_a (\rho u^a) = 0$ can be rewritten in the non-covariant, Newtonian-looking, but exact form $\partial_t \rho_* + \partial_i (\rho_* v^i) = 0$, where $v^i = u^i / u^0$ is the ordinary coordinate velocity.} Summing these expressions, and making use of Eq.~\eqref{1st_law_thermo}, the final result for the first law \eqref{bim!} in the case of a perfect fluid matter source reads
\beq\label{1st_law_PF}
	\delta H_\xi = \int_\Sigma \left[ \, \bar{\mu} \, \Delta(\ud M_\text{bar}) + \overline{T} \, \Delta(\ud S) + w^a \Delta(\ud C_a) \right] ,
\eeq
where we further introduced the redshifted temperature $\overline{T} =z T$, the redshifted (specific) chemical potential $\bar{\mu} = z (H \!-\! T s)$, the entropy element $\ud S = s \, \ud M_\text{bar}$, and the circulation element $\ud C_a = H u_a \ud M_\text{bar}$, itself defined from the current of enthalpy $H u_a$, which is related to the fluid vorticity $\omega_{ab} = \nabla_{[a} \bigl( H u_{b]} \bigr)$; see e.g. Refs.~\cite{Go.06,AnCo.07}. The result \eqref{1st_law_PF} is in complete agreement with the generalized first law of Friedman, Ury{\=u} and Shibata \cite{Fr.al.02}, in the particular case of a perfect fluid matter distribution. For black hole spacetimes, the known additional event horizon contribution could easily be recovered from the inner boundary term in the identity \eqref{identity}.

Finally, in the specific case of binary system of neutron stars moving along an exactly circular orbit, with a helical Killing field \eqref{k}, we note that the generalized first law \eqref{1st_law_PF} (with $\delta H_k = \delta M - \Omega \, \delta J$) is reminicent of a result derived long ago by Thorne, Zeldovich and Novikov \cite{Th.67,ZeNo} (see also Ref.~\cite{Ca2.73}). For a star undergoing a thermodynamically reversible process between two stationary and axisymmetric equilibrium configurations, the changes in the mass $M_\star$ and angular momentum $J_\star$ of the star are related by
\beq
    \delta M_\star - \Omega_\star \, \delta J_\star = \bar{\mu} \, \delta M_\text{bar} + \overline{T} \, \delta S \, ,
\eeq
where $\Omega_\star$ denotes the circular orbital velocity of the fluid element where the exchange of particles (or baryonic mass) and entropy takes place. No term of the form $w^a \, \delta C_a$ appears in their expression because the 4-velocity of each fluid element is assumed to be aligned with the linear combination $t^a + \Omega_\star \, \phi^a$ of the timelike and rotational Killing vectors.

\section{Komar integral}\label{app:Komar}

In this appendix, we show that the first integral relation obtained in Eq.~\eqref{1st_int_spin} can also be derived from the Komar-type conserved quantity associated to the existence of the helical Killing field \eqref{k}. We start from Eq.~\eqref{kom1} and employ the notations introduced in Sec.~\ref{sec:varid}. Using the Noether theorem, Eqs.~\eqref{J_g} and \eqref{Noether}, the Kostant formula (see Paper I) and the Einstein equation, we readily obtain
\beq\label{Komar}
    M_\text{K} - 2\Omega J_\text{K} = 2 \int_\infty \bm{Q}_\ug[k] = - 2 \int_\Sigma \, \bigl( T^{ab} k_b - \tfrac{1}{2} T k^a \bigr) \, \ud \Sigma_a \, ,
\eeq
where $T \equiv g_{ab} T^{ab}$ is the trace of the energy-momentum tensor and $\Sigma$ any spacelike hypersurface bounded by a 2-sphere at spatial infinity. This formula was previously written down in Eq. (3.18) of Ref.~\cite{Le.al.12}. It is closely related to the Tolman formula for the mass and angular momentum of a stationary-axisymmetric star \cite{Tol}.

Let $F(\Sigma)$ denote the integral in the right-hand side of the identity \eqref{Komar}. This integral can be shown to be independent of the choice of hypersurface $\Sigma$, as was done in Sec.~\ref{subsec:arbitrary} for the integrals \eqref{integrals}. Let $V$ denote a volume bounded by two spacelike hypersurfaces $\Sigma_1$ and $\Sigma_2$ and by a worldtube that includes the support of $T^{ab}$. Then by using Stokes’ theorem we readily find
\beq
    F(\Sigma_1) - F(\Sigma_2) = \int_V \nabla_a \bigl( T^{ab} k_b - \tfrac{1}{2} T k^a \bigr) \, \ud V = 0 \, ,
\eeq
as a consequence of the local conservation law $\nabla_a T^{ab} = 0$, of Killing's equation $\nabla_{(a} k_{b)} = 0$, which implies $\nabla_a k^a = 0$, and of the Lie-dragging $\Lik T = k^a \nabla_a T = 0$ along $k^a$ of the trace $T$, itself a consequence of the Lie-dragging of $T^{ab}$ (see Paper I).

To evaluate the right-hand side of Eq.~\eqref{Komar}, we note that the first term is equal to (minus twice) the integral $I$ defined in Eq.~\eqref{I}. For the second term, we choose a hypersurface $\Sigma$ that is orthogonal to $\gamma$ at the intersection point $\scP\equiv\Sigma\cap\gamma$, as in Sec.~\ref{sec:firstlaw}. We may then write explicitly the trace $T$ of the energy-momentum tensor \eqref{SET}, to obtain
\beq\label{Komar2}
    \int_\Sigma T k^a \ud \Sigma_a = - \int_V \bigl[ p^a u_a \, \delta_4 + \nabla_a (D^a\delta_4) \bigr] k^b n_b \, \ud V = -p^a u_a k^b n_b + D^a \nabla_a (k^b n_b) \, ,
\eeq
where $D^a = - S^{ab} u_b$ is the mass dipole defined in Eq.~\eqref{zip!} and $n^a$ is the unit vector normal to $\Sigma$. In the second equality we used the defining property of the invariant Dirac distribution and Stokes' theorem to integrate by parts the dipolar term (see Sec.~\ref{sec:firstlaw} for details).

This expression can be further simplified by means of the formulas \eqref{k=|k|n_bis}--\eqref{gradient} derived in Sec.~\ref{subsec:hypersurface}. Combining this with the formula \eqref{K_3} for the integral $I$ and summing over the particles, we finally obtain
\beq\label{bingo}
	M_\text{K} - 2 \Omega J_\text{K} = -\sum_\ui \bigl( p^a_\ui k_a  + S_\ui^{ab} \nabla_a k_b - D^a_\ui \dot{k}_a \bigr) \, .
\eeq
This formula can be given an alternative form by using the decomposition \eqref{zip!} of the spin tensor $S^{ab}$ in terms of the spin vector $S^a$ and the mass dipole $D^a$, as well as the definition \eqref{vorticity} of the vorticity $V^a$, yielding
\beq\label{bingo_bis}
	M_\text{K} - 2 \Omega J_\text{K} = -\sum_\ui \bigl( p^a_\ui k_a + 2 S_\ui^a V_a + D^a_\ui \dot{k}_a \bigr) \, .
\eeq

\bibliography{ListeRef.bib}

\end{document}